\documentclass[twocolumn]{aastex701} 
\usepackage{amsmath} 
\usepackage{mathtools} 

\begin{document}
\title{
Detectability of 100 TeV Gamma Rays from PeV Cosmic Rays\\
Accelerated in the Early Phase of a Supernova Shock Wave
}

\author[orcid=0009-0000-7690-5197]{Tomotaka Nishikawa}
\affiliation{Department of Physics, Graduate School of Science, Nagoya University, Furo-cho, Chikusa-ku, Nagoya 464-8602, Japan}
\email[show]{nshikawa.tomotaka.c4@s.mail.nagoya-u.ac.jp}

\author[orcid=0000-0002-7935-8771]{Tsuyoshi Inoue}
\affiliation{Department of Physics, Konan University, Okamoto 8-9-1, Higashinada-ku, Kobe 658-8501, Japan}
\email{tsuyoshi.inoue@konan-u.ac.jp}

\author[orcid=0000-0002-3971-0910]{Alexandre Marcowith}
\affiliation{Laboratoire Univers et Particules de Montpellier (LUPM), Universit\'e Montpellier, CNRS/IN2P3, CC72, place Eug\`ene Bataillon, 34095 Montpellier Cedex 5, France}
\email{almarcowith@gmail.com}

\correspondingauthor{Tomotaka Nishikawa}


\begin{abstract}
Galactic cosmic rays (CRs) are thought to be accelerated by diffusive shock acceleration in supernova remnants (SNRs), 
but observations of SNRs aged $\sim10^{2}$--$10^{3}$ yr suggest maximum CR energies below the ${\rm PeV}$ level. 
Recently, \citet{2021ApJ...922....7I} 
showed with kinetic-MHD simulations that CRs can be accelerated to $\lesssim3\,{\rm PeV}$ when a blast wave shock propagates through dense 
circumstellar material (CSM) within tens of days after explosion. 
Such CSM can be produced by a red-supergiant (RSG) wind with an observationally motivated mass-loss rate of $\sim10^{-3}\,{\rm M_{\odot}\,yr^{-1}}$. 
$100\,{\rm TeV}$ gamma-rays from neutral pion decay can test ${\rm PeV}$ CR acceleration.
However, hadronic gamma-rays from very young SNRs can be significantly 
attenuated by supernova photospheric photons and the cosmic background radiation. 
Using the CR distribution functions and CSM density profiles from \citet{2021ApJ...922....7I}, 
we calculate time-dependent gamma-ray fluxes from Type~II-P SNe, including both attenuation processes. 
We find that photospheric photons 
attenuate the intrinsic $1$--$100\,{\rm TeV}$ gamma-ray flux by about one order of magnitude. 
For the dense CSM case with $\dot{M}_{\rm RSG}\sim10^{-3}\,{\rm M_{\odot}\,yr^{-1}}$ 
and an outer CSM radius of $\sim3\times10^{15}\,{\rm cm}$, 
CTA can detect $100\, {\rm TeV}$ gamma-rays with a $50\, {\rm h}$ observation out 
to $\sim 4\,{\rm Mpc}$ (this is reduced to $\sim 3\,{\rm Mpc}$ 
if the spatial extent of the CSM is shrunk to $2 \times 10^{15}\,{\rm cm}$).
These results suggest that early $100\,{\rm TeV}$ observations of nearby Type~II-P SNe can probe PeV CR acceleration in sufficiently dense and extended CSM, 
although detectability depends on the CSM structure and occurrence fraction. 
\end{abstract}

\keywords{
\uat{Gamma-rays}{637} ---
\uat{Cosmic ray astronomy}{324} ---
\uat{Supernova remnants}{1667}
}


\section{Introduction}\label{sec:Int}
Cosmic rays (CRs) are charged particles mostly composed of nuclei that 
shower down on the Earth and were discovered by V.F. Hess in 1912. 
The observed CR spectrum spans a wide energy range from $10^8\, {\rm eV}$ to beyond $10^{20}\, {\rm eV}$. 
The CR spectrum exhibits a power law that has a characteristic break at $10^{15.5}\, {\rm eV}$ so-called the knee-energy. 
Supernova remnants (SNRs) are considered to be the most promising site of Galactic CR acceleration up to the knee energy. 
It is widely accepted that the diffusive shock acceleration (DSA) that operates around the blast wave shock of an SNR takes charge of CR acceleration 
\citep{1983RPPh...46..973D, 1987PhR...154....1B}.
It has been believed that SNRs in the late free-expansion phase or early Sedov–Taylor phase can confine CRs up to the knee energy. 
However, gamma-ray observations of SNRs with ages $\sim 10^{2}-10^{3}$ years indicate 
that the maximum energy of CRs is around $20\, {\rm TeV}$ 
\citep{2012ApJ...744L...2G, 2013APh....43...71A, 2017MNRAS.472.2956A, 2018A&A...612A...5H, 2022ApJ...924...45S}. 
Therefore, it is argued that younger SNRs in which the shock wave propagating in a circumstellar medium (CSM) can be more advantageous for acceleration up to the knee-energy 
\citep{2013MNRAS.431..415B, 2013MNRAS.435.1174S, 2018MNRAS.479.4470M}. 
Assuming a high-density CSM created by the wind of a progenitor red supergiant (RSG) of the supernova (SN), CRs themselves can induce the non-resonant hybrid instability (NRHI) (or Bell instability) to amplify the magnetic field in the upstream region. 
\citet{2018MNRAS.479.4470M}
discussed that the DSA can achieve acceleration up to the knee energy in such an environment within a few tens of days after the explosion.

Gamma-rays in the $\sim 100\, {\rm TeV}$ range are generated by collisions between $\rm PeV$ CRs and the CSM, 
but several radiative processes attenuate the gamma-ray flux. 
Among the attenuation effects, the dominant one is the photon-photon annihilation process between gamma-rays and photospheric photons from the SN photosphere 
\citep{PhysRevLett.16.252, PhysRev.155.1408}. 
This effect has been discussed in high-mass X-ray binaries composed of neutron stars or black holes orbiting early-type stars 
\citep{2006A&A...451....9D} 
and in core-collapse supernovae (CCSNe) 
\citep{2009A&A...499..191T, 2014NuPhS.256...94M}. 

Recently, 
\citet{2020MNRAS.494.2760C} 
computed the optical depth of gamma-rays from the very early phase of 
the Type~IIb SN~1993J 
by taking into account the detailed geometry of the forward shock and the SN photosphere. 
A follow-up study by 
\citet{2022MNRAS.511.3321C} 
extended this calculation to Type~II-P SNe and showed that 
pair production with soft photospheric photons can severely attenuate 
the TeV gamma-ray flux during the first days to weeks after the explosion. 
They concluded that most extragalactic Type~II-P events would not be detectable 
even with a $50\,{\rm h}$ observation by the Cherenkov Telescope Array 
\citep[CTA;][]{2019scta.book.....C}. 
In Type~II-P SNe, this attenuation can last longer because the forward shock 
remains relatively close to the photosphere, thereby severely limiting 
the detectability horizon for TeV gamma-rays.

In the DSA process, CRs gain energy by scattering off magnetic turbulence. 
There are various instabilities that can induce magnetic turbulence 
\citep{2016RPPh...79d6901M}. 
The NRHI is predicted to be the most important instability for the acceleration of CRs \citep{2004MNRAS.353..550B}. 
Recently, by employing a novel numerical technique, it has become possible to compute CR acceleration simultaneously with the evolution of the NRHI under realistic conditions 
\citep{2021ApJ...922....7I, 2024ApJ...965..113I}. 
It has been shown that the NRHI does indeed amplify the magnetic field to a level consistent with observations. 

Recent early-time observations of Type~II SNe suggest that some RSG progenitors experience enhanced mass loss shortly before explosion, with inferred mass-loss rates up to about two orders of magnitude higher than the conventional value of $\dot{M}_{\rm RSG} \sim 10^{-5}\, {\rm M_\odot\, yr^{-1}}$
\citep{2013Natur.494...65O, 2017NatPh..13..510Y, 2018NatAs...2..808F}. 
\citet{2021ApJ...922....7I} calculated particle acceleration in such a dense CSM around RSGs, showing that the maximum energy reaches the knee within a few tens of days after the explosion. 
In such a dense CSM environment, hadronic gamma-rays are naturally expected to be brighter than in the case of the conventional mass-loss rate, making it important to re-examine the gamma-ray detectability. 
Thus, in this paper, we re-evaluate the gamma-ray flux emitted from very young SNR surrounded by the dense CSM with such enhanced mass-loss rates by employing the CR acceleration simulation data of 
\citet{2021ApJ...922....7I}. 

The structure of this paper is as follows. 
In Section~\ref{sec:Method-GF}, as preparation for predicting gamma-ray observations, we present the numerical setup and results of the CR acceleration simulations conducted by 
\citet{2021ApJ...922....7I}. 
In addition, we describe the computational method used to account for the attenuation of gamma-rays due to pair production with ambient photons. 
In Section~\ref{sec:Result}, we present the numerical results and their physical interpretation. 
In Section~\ref{Discussion}, we evaluate the maximum detectable distance for CTA observations 
and estimate the corresponding event rates. 
Finally, in Section~\ref{Summary}, we summarize the findings of this study.

\section{Method: Gamma-ray flux} \label{sec:Method-GF}
\subsection{Circumstellar medium and cosmic ray profile} \label{sec:CR&CSM}
In this section, we summarize the isotropic component of the CR distribution function, $f(r,p)$, and the CSM density profile, which are used to evaluate the gamma-ray flux.
These quantities are based on recalculated CR distribution functions and CSM density profiles from the setup of \citet{2021ApJ...922....7I}, using the formulation adopted in \citet{2024ApJ...965..113I}.
The adopted simulations assume a RSG CSM model following \citet{2018MNRAS.479.4470M}.
The CSM mass density is characterized by the progenitor mass-loss rate $\dot{M}_{\rm RSG}$ and the wind velocity $v_{\rm wind}$ as
\begin{align}
    \rho_{\rm CSM}(r)
    &=
    5\times10^{-15}\, {\rm g\,cm^{-3}}
    \left(\frac{r}{10^{14}\,{\rm cm}}\right)^{-2}\notag \\
    &\times
    \left(\frac{\dot{M}_{\rm RSG}}{10^{-5}\,{\rm M_{\odot}\,yr^{-1}}}\right)
    \left(\frac{v_{\rm wind}}{10\,{\rm km\,s^{-1}}}\right)^{-1}.
    \label{eq:rho_CSM}
\end{align}
In these simulations, a hybrid system of the Bell-MHD equations and the telegrapher-type diffusion--convection equations for CRs is solved in polar coordinates, following \citet{2021ApJ...922....7I} and \citet{2024ApJ...965..113I}. 
This formulation evolves the background plasma together with the isotropic and first-order anisotropic components of the CR distribution function, and follows the coupling among CR acceleration, CR streaming, the induced return current, and magnetic-field amplification by the NRHI. 
In this framework, CRs propagating into the shock upstream region are included in the radial profile of the CR distribution function, $f(r,p)$. 
At the epochs adopted in this work, the amplified magnetic turbulence still confines most CRs in the vicinity of the forward shock. 

The adopted spatial resolution is sufficient to resolve the growth of the NRHI and to follow its saturation through the advection of the amplified magnetic turbulence at the forward shock.
The mean molecular weight of the CSM is set to $1.27\,m_{\rm p}$, so that the CSM number density is given by $n(r)=\rho_{\rm CSM}(r)/(1.27\,m_{\rm p})$.
The CRs are treated as pure protons, where $m_{\rm p}$ is the proton mass.
In particular, among the models considered in \citet{2021ApJ...922....7I}, we adopt the results of Model $0$ and Model $1$.

Model $0$ uses the conventional RSG mass-loss rate $\dot{M}_{\rm RSG}=10^{-5}\,{\rm M_{\odot}\,yr^{-1}}$, 
while Model $1$ adopts the modern, observationally motivated RSG mass-loss rate 
$\dot{M}_{\rm RSG}=10^{-3}\,{\rm M_{\odot}\,yr^{-1}}$. 
Other key parameters are common to both models: wind velocity $v_{\rm wind}=10\,{\rm km\,s^{-1}}$, cosmic-ray injection fraction at the shock $\eta_{\rm inj}=6\times10^{-4}$, and ejecta velocity $v_{\rm ej}=10^{4}\,{\rm km\,s^{-1}}$. 
Here, $\eta_{\rm inj}$ denotes the fraction of thermal gas particles injected into the acceleration process.
In the present gamma-ray calculation, the CR acceleration efficiency is not prescribed as an independent free parameter to normalize the gamma-ray flux.
Instead, the magnitude and spectral shape of the CR distribution function are taken from the kinetic-MHD simulation results obtained with the adopted injection fraction, $\eta_{\rm inj}=6\times10^{-4}$. 
Therefore, resulting gamma-ray flux depend on the adopted value of $\eta_{\rm inj}$. 
A parameter survey with respect to $\eta_{\rm inj}$ is beyond the scope of this work, and this dependence remains one of the uncertainties in the present calculation. 
We discuss the effective CR acceleration efficiency corresponding to this adopted CR distribution function in Appendix~\ref{sec:E_CR_tot}.
For the present study, we use CR distribution functions and CSM density profiles that were recalculated from the setup of \citet{2021ApJ...922....7I} using the formulation adopted in \citet{2024ApJ...965..113I}.
The recalculated CR distribution functions and CSM density profiles are qualitatively similar to those presented in \citet{2021ApJ...922....7I}; however, the maximum CR energy in Model $1$ at the end of the simulation is slightly higher, increasing from $1.8\,{\rm PeV}$ to $3.2\,{\rm PeV}$.

\begin{figure}[tbp]
    \centering
    \includegraphics[width = \linewidth, height=0.6\textheight, keepaspectratio]{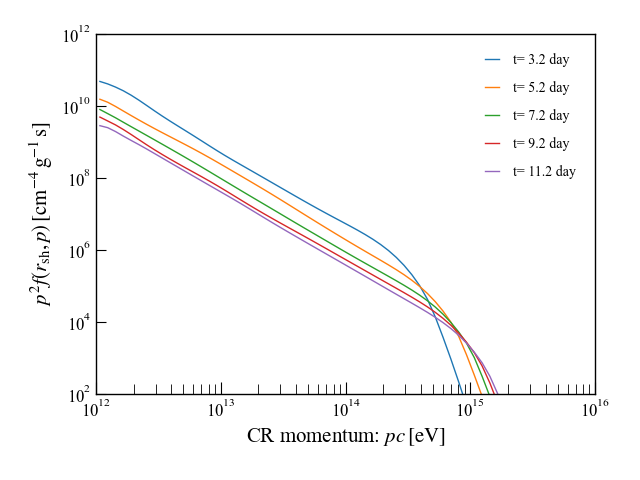}
    \caption{
    CR distribution function multiplied by $p^2$ near the shock front, 
    calculated for Model $0$ of \citet{2021ApJ...922....7I}. 
    Model $0$ adopts a conventional RSG wind with 
    $\dot{M}_{\rm RSG}=10^{-5}\,{\rm M_\odot\,yr^{-1}}$, 
    $v_{\rm wind}=10\,{\rm km\,s^{-1}}$, and 
    $\eta_{\rm inj}=6\times10^{-4}$. 
    The colors of the solid lines indicate the elapsed time since core-collapse event. 
    Each curve is obtained by spatially averaging the CR distribution function $f(r,p)$ 
    over the radial range from $r_{\rm Sh}-50\,\Delta r$ to $r_{\rm Sh}$, 
    where $r_{\rm Sh}$ is the shock-front radius and 
    $\Delta r = 9.5 \times 10^{8}\,{\rm cm}$ 
    is the numerical resolution. 
    }
    \label{energy-f_CR_LML}
\end{figure}

\begin{figure}[tbp]
    \centering
    \includegraphics[width = \linewidth, height=0.6\textheight, keepaspectratio]{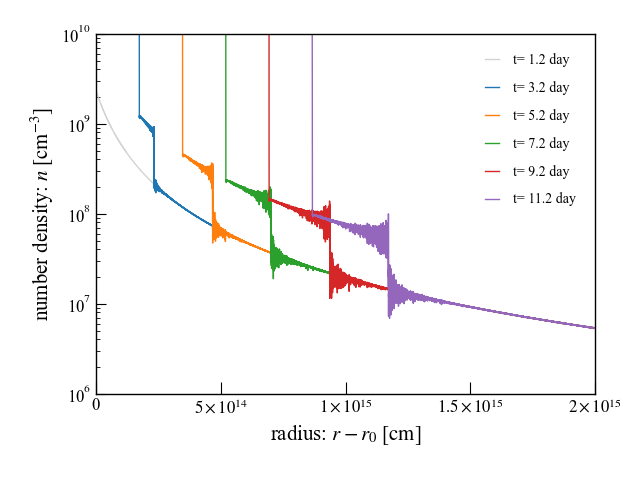}
    \caption{
    Radial density profile of the background fluid calculated for Model $0$ of 
    \citet{2021ApJ...922....7I}. 
    Model $0$ adopts a conventional RSG wind with 
    $\dot{M}_{\rm RSG}=10^{-5}\,{\rm M_\odot\,yr^{-1}}$ and 
    $v_{\rm wind}=10\,{\rm km\,s^{-1}}$. 
    The colors of the solid lines indicate the elapsed time since core-collapse event. 
    The snapshots are shown every $2\,{\rm days}$ over a period of $10\,{\rm days}$, 
    starting from $t=1.2\,{\rm days}$, which corresponds to the time when the ejecta reaches 
    $10^{14}\,{\rm cm}$ from the stellar center.
    }
    \label{r-n_CSM_LML}
\end{figure}

\begin{figure}[tbp]
    \centering
    \includegraphics[width = \linewidth, height=0.6\textheight, keepaspectratio]{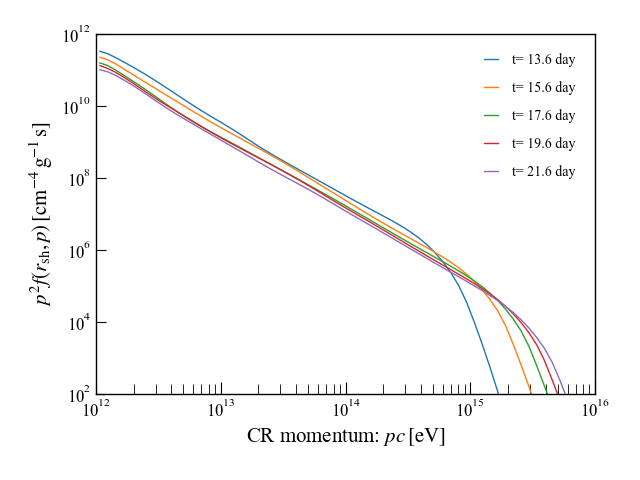}
    \caption{
    Same as Figure~\ref{energy-f_CR_LML}, but for Model $1$ of 
    \citet{2021ApJ...922....7I}. 
    Model $1$ adopts an enhanced RSG wind with 
    $\dot{M}_{\rm RSG}=10^{-3}\,{\rm M_\odot\,yr^{-1}}$, 
    $v_{\rm wind}=10\,{\rm km\,s^{-1}}$, and 
    $\eta_{\rm inj}=6\times10^{-4}$.
    }
    \label{energy-f_CR}
\end{figure}

\begin{figure}[tbp]
    \centering
    \includegraphics[width = \linewidth, height=0.6\textheight, keepaspectratio]{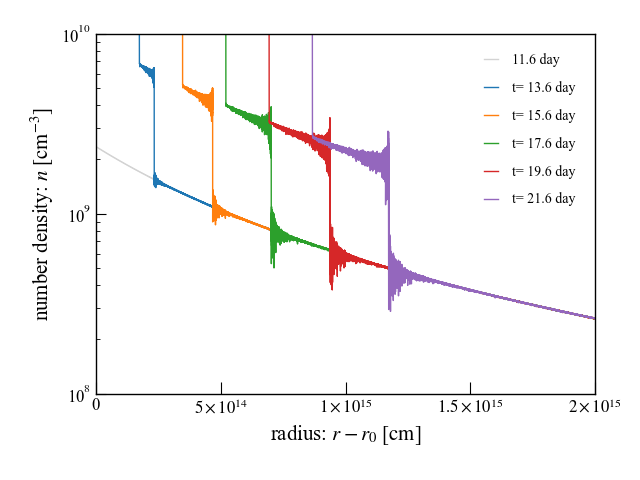}
    \caption{
    Same as Figure~\ref{r-n_CSM_LML}, but for Model $1$ of 
    \citet{2021ApJ...922....7I}. 
    Model $1$ adopts an enhanced RSG wind with 
    $\dot{M}_{\rm RSG}=10^{-3}\,{\rm M_\odot\,yr^{-1}}$ and 
    $v_{\rm wind}=10\,{\rm km\,s^{-1}}$. 
    The snapshots are shown every $2\,{\rm days}$ over a period of $10\,{\rm days}$, 
    starting from approximately $11.6\,{\rm days}$ after the core-collapse event, 
    which corresponds to the time when the ejecta reaches 
    $10^{15}\,{\rm cm}$ from the stellar center.
    }
    \label{r-n_CSM}
\end{figure}

Figures~\ref{energy-f_CR_LML} and \ref{r-n_CSM_LML} show the cosmic-ray spectra near the shock front and the density structure of the background fluid for Model $0$, while Figures~\ref{energy-f_CR} and \ref{r-n_CSM} present those for Model $1$. 
Colors represent the time elapsed since core-collapse. 
The density profiles shown in Figures~\ref{r-n_CSM_LML} and \ref{r-n_CSM} exhibit small-scale oscillatory structures near the shock front.
As discussed by \citet{2021ApJ...922....7I}, these fluctuations are attributed to the back-reaction of Alfv\'en waves induced by the NRHI. 
Note that 
\citet{2021ApJ...922....7I} defined the time $t=0$ as the moment when the shock front passes $r_0 = 10^{14}\, {\rm cm}$ from the stellar center for Model $0$ and $r_0 = 10^{15}\, {\rm cm}$ for Model $1$. 
However, in this paper we define $t=0$ as the time of the core-collapse. 
Thus, the initial conditions for Model $0$ and Model $1$ respectively correspond to $1.2\, {\rm days}$ and $11.6\, {\rm days}$ after the core-collapse. 
Note also that, in such an early phase of shock propagation, 
the effect of free--free cooling can be important, which potentially suppresses CR acceleration \citep{2025ApJ...984..103K}. 
However, in our numerical models, because the simulations start in the somewhat outer part of the CSM, 
the densities are already low enough for free--free cooling to be ineffective. 
For instance, in Model $0$ and Model $1$, the ratios of the cooling time due to free--free emission 
\citep{1979rpa..book.....R} 
to the dynamical time $t$ are calculated to be $29.35$ and $2.94$, respectively, at the starting point of the simulations, 
which justifies the neglect of free--free cooling in the simulations. 
Nevertheless, this does not imply that CR acceleration is always possible in dense CSM environments such as our Model $1$, because even a modest enhancement of the CSM density by a factor of $\sim 3$ makes free--free cooling effective and suppresses CR acceleration. 
The spatial extent of the dense CSM is also an important factor in determining the suppression of CR acceleration. 
Model $1$ implicitly assumes that the dense CSM extends to an outer radius of $\sim 3\times10^{15}\,{\rm cm}$, which may not be typical for many Type~II-P SN environments.

\subsection{Intrinsic gamma-ray flux} \label{sec:UnattGF}
In this section, we estimate the intrinsic gamma-ray flux, i.e., 
the flux before attenuation by surrounding soft photons, based on 
\citet{1971NASSP.249.....S}. 
The gamma-ray production rate for energies above $E_{\gamma}$,
per unit volume and per unit time, at a radius $r$ in the SNR,
$Q(r, E \ge E_{\gamma})$, is given by
\begin{align}
    Q(r, E \ge E_{\gamma}) 
    &= 
    n(r)
    \int_{E_{\gamma}}^{+\infty}
    {\rm d}E'
    \int_{0}^{+\infty}
    {\rm d}E_{\rm p}\, \notag \\
    \times &N_{\rm p}(r, E_{\rm p})
    v_{\rm p}
    \frac{{\rm d}\sigma_{\rm pp \rightarrow \pi^{0}}(E_{\rm p}, E')}{ {\rm d}E' }, 
\end{align}
where $E$ (or $E'$) denotes the energy of a gamma-ray photon. 
$n(r)$ denotes the number density of protons in the background medium at radius $r$, 
while $E_{\rm p}$, $v_{\rm p}$, and $N_{\rm p}(r, E_{\rm p})$ represent the energy, velocity, and differential number density per unit energy of CR protons, respectively. 
Here, $n(r)$ is obtained from the CSM density profiles shown in Figures~\ref{r-n_CSM_LML} and \ref{r-n_CSM}. 

The CR proton density $N_{\rm p}(r,E_{\rm p})$ is calculated from the isotropic component of the CR distribution function $f(r,p)$ shown in Figures~\ref{energy-f_CR_LML} and \ref{energy-f_CR} as
\begin{align}
    N_{\rm p}(r,E_{\rm p})\,{\rm d}E_{\rm p}
    =
    4\pi p^2 f(r,p)\,{\rm d}p .
\end{align}
The cross section for neutral pion production in proton--proton collisions, 
$\sigma_{\rm pp \rightarrow \pi^{0}}$, is taken from the parametric model fitted to experimental data presented in 
\citet{2007ApJ...662..779K}. 
The intrinsic gamma-ray flux at a distance $D$, 
$F_{\gamma, {\rm intr}}(E \ge E_{\gamma})$, 
is obtained by integrating the gamma-ray production rate 
$Q(r, E \ge E_{\gamma})$ over the region where the accelerated CRs are distributed around the shock front:
\begin{align}
    F_{\gamma, {\rm intr}} ( E \ge E_{\gamma} )
    &=
    \frac
    {\int Q(r, E \ge E_{\gamma}) \, {\rm d}^3 r}
    {4 \pi D^2}.
\end{align}
Because $Q(r,E \ge E_{\gamma})$ is evaluated using the spatially dependent CR distribution function and CSM density profile, this integration includes the contribution from CRs distributed around the shock front.

\subsection{Evolution of the SN shock and photosphere} \label{sec:SF&PSmodel}
\begin{figure}[tbp]
    \centering
    \includegraphics[width = \linewidth, height=0.6\textheight, keepaspectratio]{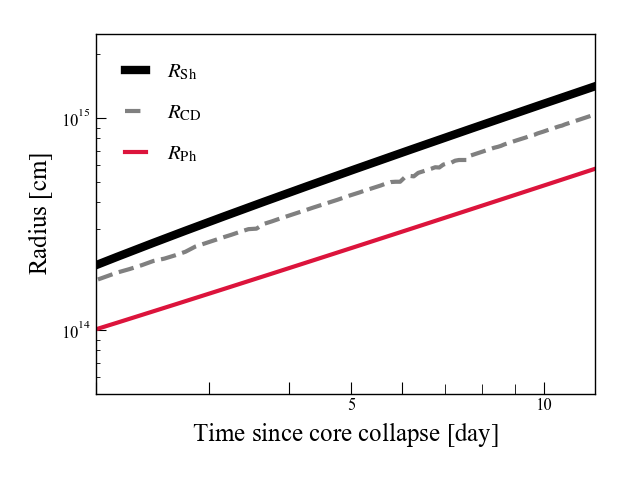}
    \caption{
    Time evolution of the shock front radius $R_{\rm Sh}$ (black thick line), 
    the contact discontinuity radius $R_{\rm CD}$ (gray thin dashed line), 
    and the photospheric radius $R_{\rm Ph}$ (red thin line) 
    for an outer ejecta velocity 
    $v_{\rm ej} = 10^{4}\, {\rm km\, s^{-1}}$ 
    and an ejecta mass of $M_{\rm ej} = 2.67\, {\rm M_{\odot}}$. 
    The values of $R_{\rm Sh}$ and $R_{\rm CD}$ are taken from the simulation results of Model $0$ in 
    \citet{2021ApJ...922....7I}, 
    while $R_{\rm Ph}$ is based on the analytical model of 
    \citet{2018ApJ...868L..24L}.
    }
    \label{Time-R_Ph_CD_sh_0}
\end{figure}

\begin{figure}[tbp]
    \centering
    \includegraphics[width = \linewidth, height=0.6\textheight, keepaspectratio]{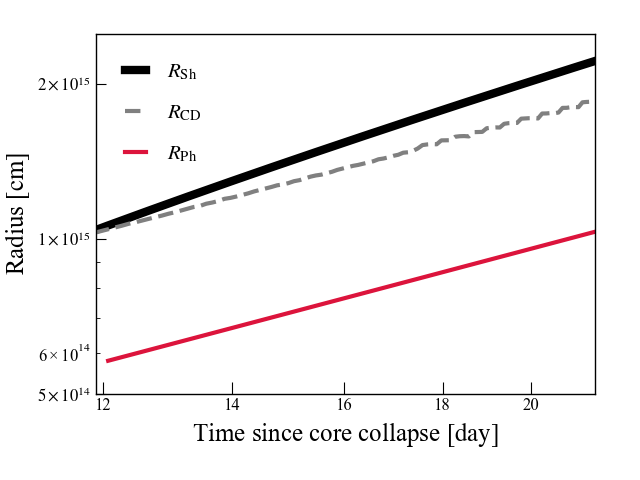}
    \caption{
    Same as Figure~\ref{Time-R_Ph_CD_sh_0}, but for Model $1$ in \citet{2021ApJ...922....7I}.
    }
    \label{Time-R_Ph_CD_sh_1}
\end{figure}

In this section, we describe the evolutions of the SN shock and photosphere, 
focusing on the photospheric radius and surface temperature required for calculating the gamma-ray opacity. 
In this calculation, we consider Type II-P SNe, which account for approximately $60\, \%$ of CCSNe 
\citep{2009MNRAS.395.1409S}. 
For the evolution of the photospheric radius, we use the model of 
\citet{2018ApJ...868L..24L}. 
Figures~\ref{Time-R_Ph_CD_sh_0} and \ref{Time-R_Ph_CD_sh_1} show the evolution of the shock front radius $R_{\rm Sh}$ and the contact discontinuity radius $R_{\rm CD}$ obtained from Model $0$ and Model $1$ by \citet{2021ApJ...922....7I}, 
as well as the photospheric radius $R_{\rm Ph}$ obtained from the analytical model of \citet{2018ApJ...868L..24L}. 
In calculating the photospheric radius, we adopt an ejecta mass of $M_{\rm ej} = 2.67\, {\rm M_{\odot}}$ to ensure consistency with the outer ejecta velocity 
$v_{\rm ej} = 10^{4}\, {\rm km\, s^{-1}}$ assumed in the cosmic-ray acceleration simulations of 
\citet{2021ApJ...922....7I}.
The photospheric temperature is modeled using the analytical model of 
\citet{2011ApJ...728...63R} and is given by
\begin{align}
    T_{\rm Ph}(t)
    &=
    1.7\, {\rm eV}\,
    \left( \frac{f_{\rm \rho}}{0.1} \right)^{-0.037}
    \left( \frac{E_{\rm SN}}{10^{51}\,{\rm erg}} \right)^{0.027}
    \left( \frac{R_{*}}{10^{13}\, {\rm cm}} \right)^{1/4} \notag \\
    & \times \left( \frac{\kappa}{0.34\, {\rm cm^{2}\, g^{-1}}} \right)^{-0.28} 
    \left( \frac{M_{\rm ej}}{\rm M_{\odot}} \right)^{-0.054}
    \left( \frac{t}{10^5\,{\rm s}} \right)^{-0.45}. 
    \label{T_ph_type2}
\end{align}\\
Here, $f_{\rho}$ is a parameter that depends on the density structure of the ejecta. 
For an RSG progenitor, $f_{\rho}$ takes values in the range 
$0.079 \lesssim f_{\rho} \lesssim 0.13$ 
\citep{2004MNRAS.351..694C}, 
and we adopt $f_{\rho} = 0.1$.
$E_{\rm SN}$ is the kinetic energy of the SN explosion, 
$R_{*}$ is the radius of the progenitor star before the explosion, 
$\kappa$ is the opacity assumed to be constant in time and space, 
$M_{\rm ej}$ is the ejecta mass, 
and $t$ is the elapsed time since core-collapse. 
We assume that the SN photosphere emits blackbody radiation. 
The differential photon number density $n_{\rm Ph}(\epsilon, T_{\rm Ph}(t))$, 
where $\epsilon$ is the soft photon energy, 
is defined as the number of soft photons 
per unit volume, per unit photon energy, 
and per unit solid angle: 
\begin{align}
    n_{\rm Ph}(\epsilon, T_{\rm Ph}(t))
    &=
    \frac{2 \epsilon^2}{h^3c^3}
    \frac{1}{\exp \left( \epsilon/ k_{\rm B} T_{\rm Ph}(t) \right)-1} \, \notag \\
    &\left[ {\rm cm^{-3}\, erg^{-1}\, sr^{-1}} \right]. 
    \label{eq:n_BB}
\end{align}

\subsection{Gamma-ray flux attenuation} \label{sec:AttGF}
When photons pass through the interstellar medium (ISM) and the CSM, 
they undergo absorption and scattering through various processes. 
For high energy photons, such as gamma-rays, 
the photon--photon pair production process ($\gamma\gamma \rightarrow e^{+}e^{-}$) is dominant. 
To quantitatively express the attenuation of the gamma-ray flux, 
we compute the gamma-ray opacity based on 
\citet{2020MNRAS.494.2760C}. 
Gamma-rays produced near the shock front of a SNR interact with soft photons from the SN photosphere, 
as well as with the cosmic background radiation (CBR), 
during their propagation toward Earth. 
Here, gamma-rays also interact with photons originating from the thermal bremsstrahlung of the shocked CSM. 
However, their contribution to the opacity is sufficiently small compared to those from the other two sources 
(see Appendix~\ref{sec:shockedCSM}). 

\begin{figure}[tbp]
    \centering
    \includegraphics[width = \linewidth, height=0.6\textheight, keepaspectratio]{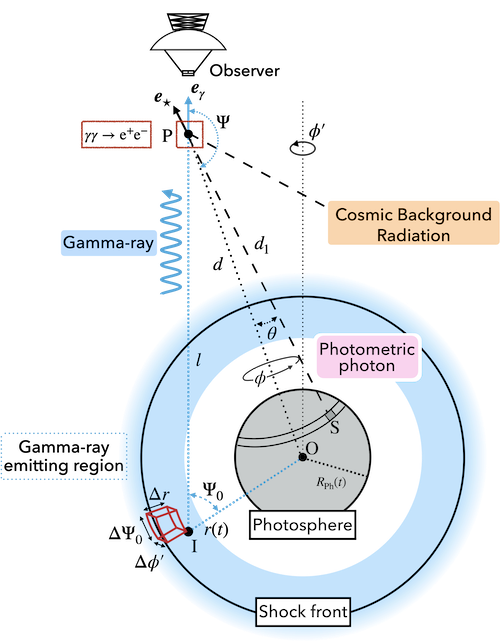}
    \caption{
    Geometry of gamma-ray propagation and attenuation in a SNR at time $t$ after the core-collapse event.
    The inner black circle represents the SN photosphere with radius $R_{\rm Ph}(t)$, and the outer black circle represents the shock front.
    The blue-shaded region around the shock front indicates the region where CRs are distributed and gamma-rays are produced.
    A gamma-ray emitted at point ${\rm I}$, located at radius $r(t)$ from the stellar center, propagates outward and interacts at point ${\rm P}$ with a photospheric photon emitted from point ${\rm S}$ on the photosphere.
    The points ${\rm I}$, ${\rm P}$, and ${\rm S}$ denote the gamma-ray emission point, the interaction point, and the photospheric photon emission point, respectively.
    }
    \label{Gamma-ray_process_SNR}
\end{figure}

We adopt the geometry illustrated in Figure~\ref{Gamma-ray_process_SNR} 
that represents the SNR at time $t$ since the core-collapse. 
The center of the remnant is marked as point ${\rm O}$, 
and the outer black solid circle centered at this point represents the shock front, 
while the inner black solid circle represents the photosphere. 
The CRs are distributed in the blue-shaded region, i.e., from the contact discontinuity to slightly upstream of the shock front. 
Thus, the hadronic gamma-rays produced by collisions between CRs and the CSM are mostly emitted from the region between the contact discontinuity and the shock front, 
with only a small fraction being emitted from upstream of the shock. 
In Figure~\ref{Gamma-ray_process_SNR}, gamma-rays emitted from point ${\rm I}$ within the blue-shaded region interact at point ${\rm P}$ on the line of sight 
with soft photons emitted from the photosphere at point ${\rm S}$. 
The azimuthal angle around the line of sight is denoted by $\phi'$, 
and the angle $\angle \mathrm{OIP}$ is denoted by $\Psi_{0}$.

The opacity of gamma-rays with energy $E$ emitted from this small region is expressed using the photon--photon pair-production cross section $\sigma_{\gamma \gamma}$ as
\begin{align}
     \tau_{\rm Ph}\left( E,\, t,\, r,\, \Psi_{0} \right)
     &= 
     \int_{0}^{+\infty} {\rm d}l\,
     \int_{c_{\rm min} (t)}^{1} {\rm d} \cos \theta \,
     \int_{0}^{2 \pi} {\rm d}\phi \,
     \int_{\epsilon_{\rm min} (E)}^{+\infty} {\rm d}\epsilon \, \notag \\
     \times \left( 1- \boldsymbol{e}_{\gamma} \cdot \boldsymbol{e}_{\star} \right)
     &n_{\rm Ph}(\epsilon, T_{\rm Ph}(t))\, 
     \sigma_{\gamma \gamma}(E,\, r,\,\Psi_{0},\, \theta,\, \phi,\, \epsilon),
     \label{tau_ph}
\end{align} 
where $l$ is the distance between points ${\rm I}$ and ${\rm P}$,
$\theta$ is the angle $\angle \mathrm{SPO}$, 
$\phi$ is the azimuthal angle around the line of sight, 
and $\epsilon$ is the energy of the soft photon. 
$\boldsymbol{e}_{\gamma}$ and $\boldsymbol{e}_{\star}$ represent the unit vectors of the propagation directions of the gamma-ray and the photospheric photon, respectively, and the angle between $\overline{\rm OP}$ and $\boldsymbol{e}_{\gamma}$ is denoted by $\Psi$. 
The factor $\left( 1 - \boldsymbol{e}_{\gamma} \cdot \boldsymbol{e}_{\star} \right)$ in the integrand of Equation~\eqref{tau_ph} 
shows that the pair-production probability depends on the angle between their propagation directions. 
This factor can be explicitly written as
\begin{align}
    1 - \boldsymbol{e}_{\gamma} \cdot \boldsymbol{e}_{\star}
    &=
    1 + \cos{\Psi} \cos{\theta} + \sin{\Psi} \cos{\phi} \sin{\theta}. 
    \label{photon_direction}
\end{align}
When $\Psi$ is expressed in terms of $\Psi_{0}$, it becomes 
\begin{align}
    \Psi
    &=
    \tan^{-1} \left( \frac{r \sin{\Psi_0}}{r \cos{\Psi_0} -l} \right) \,\,\, {\rm for} \,l \le r \cos{\Psi_{0}}, 
\end{align}
and
\begin{align}
    \Psi
    &=
    \pi+
    \tan^{-1} \left( \frac{r \sin{\Psi_0}}{r \cos{\Psi_0} -l} \right) \,\,\, {\rm for} \,l > r \cos{\Psi_{0}}. 
\end{align}
$n_{\rm Ph}$ denotes the number density of soft photons at the interaction point ${\rm P}$. 
$\sigma_{\gamma \gamma}$ is the total cross section for the photon--photon pair-production process \citep{PhysRevLett.16.252, PhysRev.155.1408}. 
This cross section depends on the angle between the propagation directions of the gamma-ray photon and the photospheric photon, 
as well as on their energies, and is given by\\
\begin{align}
    \sigma_{\gamma \gamma}
    =
    \frac{1}{2} \pi r_0^2
    \left( 1- \beta^2 \right)
    \left[
    \left(3-\beta^4 \right) \ln{\frac{1+\beta}{1-\beta}}
    -2 \beta \left( 2- \beta^2\right)
    \right].
\end{align}
Here, $r_0$ is the classical electron radius defined as $r_0 = e^{2}/(m_{\rm e} c^2)$. 
$\beta c$ denotes the velocity of the electron (and positron) in the center-of-mass frame, 
and $\beta$ is given by
\begin{align}
    \beta
    &=
    \left( 1-\frac{1}{s} \right)^{1/2},\\
    s
    &=
    \frac{\epsilon E}{2 m_e^2 c^4}
    \left( 1- \boldsymbol{e}_{\gamma} \cdot \boldsymbol{e}_{\star} \right).
\end{align}

We evaluate the emission time $t_{\rm S}$ of the photospheric photons that interact with the gamma-rays emitted from point ${\rm I}$ at time $t$. 
\citet{2020MNRAS.494.2760C} argued that this time difference has a significant impact on gamma-ray attenuation.  
The following calculations are carried out to incorporate this effect.  
The emission time $t_{\rm S}$ of the photospheric photon emitted from point ${\rm S}$, which interacts at point ${\rm P}$ with the gamma-ray emitted from point ${\rm I}$ at time $t$, 
is expressed using the distance $d_{1}$ between points ${\rm S}$ and ${\rm P}$ as
\begin{align}
    t_{\rm S}
    &= t + \frac{l}{c} - \frac{d_1}{c}. 
\end{align}
The distance $d_{1}$ is written as 
\begin{align}
    d_1
    &= d \cos \theta - \sqrt{\left( d \cos \theta \right)^2 - d^2 + R_{\rm Ph}^2(t_{\rm S})}. 
\end{align}
The distance $d$ can be expressed in terms of $l$ as
\begin{align}
    d
    &=
    \sqrt{ r^2 + l^2 -2 r l \cos{\Psi_0} }. 
\end{align}

Next, we describe the integration range for the opacity calculation in Equation~\eqref{tau_ph}.  
There exists a lower limit for the energy of soft photons that can undergo pair production with a gamma-ray of energy $E$, given by 
\begin{align}
    \epsilon_{\rm min}
    &=
    \frac{2 m_{\rm e}^2 c^4}{\left( 1- \boldsymbol{e}_{\gamma} \cdot \boldsymbol{e}_{\star} \right)\,E} . 
    \label{eq:eps_min}
\end{align}
The integration range for $\theta$ corresponds to the angular range on the photospheric surface from which soft photons can reach point ${\rm P}$. 
The maximum value of $\theta$, or equivalently the minimum value of $\cos \theta$, denoted as $c_{\rm min}$, is given by
\begin{align}
    c_{\rm min} 
    &\coloneqq (\cos \theta)_{\rm min}\notag \\
    &=
    \sqrt{1-\left( \frac {R_{\rm Ph} (t_{S})} {d}\right)^2} . 
\end{align}

Analogous to the expression for the opacity of gamma-rays due to photospheric photons, 
the opacity due to the CBR is given by
 \begin{align}
    \tau_{\rm CBR}\left( E,\, t,\, r,\, \Psi_{0}\right)
    &= 
    \int_{0}^{+\infty} {\rm d}l\,
    \int_{-1}^{1} {\rm d} \cos \theta \,
    \int_{0}^{2 \pi} {\rm d}\phi \,
    \int_{\epsilon_{\rm min}(E)}^{+\infty} {\rm d}\epsilon \notag \\
    \times \left( 1 - \boldsymbol{e}_{\gamma} \cdot \boldsymbol{e}_{\star} \right)\,
    &n_{\rm CBR}(\epsilon) \, \sigma_{\gamma \gamma} (E,\, r,\,\Psi_{0},\, \theta,\, \phi,\, \epsilon)
    . 
    \label{tau_CBR} 
 \end{align}
Here, $n_{\rm CBR}$ denotes the number density of CBR photons, 
which is assumed to be isotropic 
\citep{2000PhLB..493....1P}.

Using the opacities due to photospheric photons and the CBR, 
the net gamma-ray flux reaching Earth can be expressed as
\begin{align}
    F_{\gamma, {\rm abs}}(E \ge E_{\gamma})
    &= 
    \frac{1}{4 \pi D^{2}}
    \int_{0}^{2 \pi} {\rm d} \phi'
    \int_{\Psi_{0,{\rm min}}}^{\pi} \sin \Psi_{0}\, {\rm d}\Psi_{0} 
    \int r^2\, {\rm d}r \notag \\
    &Q(r, E \ge E_{\gamma}) 
    \exp \left( - \tau_{\rm Ph} - \tau_{\rm CBR} \right). 
    \label{net_flux_gamma-ray}
\end{align}
Here, $D$ denotes the distance to the SNR, and the integration range of $\Psi_{0}$ 
corresponds to the angular region from which gamma-rays can reach the observer 
without being blocked by the photosphere.
The minimum value of the emission angle, $\Psi_{0,{\rm min}}$, is given by 
\begin{align}
    \Psi_{0,{\rm min}}
    &= 
    \arcsin{\left( \frac{R_{\rm Ph} (t + t_{+})}{r(t)} \right)}. 
\end{align}
Here, $t+t_{+}$ denotes the time at which the gamma-ray emitted at time $t$ passes closest to the photosphere.  
Additionally, by defining $\mu \coloneq \cos \left( \Psi_{0,{\rm min}} \right)$ and considering the geometry between the gamma-ray emission point and the photosphere, we obtain
\begin{align}
    r^2
    \left( 1- \mu^2 \right)
    &=
    \left\{ R_{\rm Ph} \left( t + \frac{r \mu}{c} \right) \right\}^2. 
\end{align}
In the numerical calculation, $\mu$ is varied from $1$ to $0$, and by searching for the maximum value of $\mu$ that satisfies the above equation, $\Psi_{0,{\rm min}}$ is determined.  
Finally, performing the integral in Equation~\eqref{net_flux_gamma-ray}, 
the hadronic gamma-ray flux above $E_{\gamma}$ is evaluated.

\section{Result: Time-dependent gamma-ray flux}\label{sec:Result}
In this section, we present the time evolution of the integrated gamma-ray flux from a Type II-P SN at $D=1\,{\rm Mpc}$. 
The detectability as a function of distance is discussed in the next section. 
We consider six cases, defined by two mass-loss rate models and three gamma-ray energy thresholds. 
The parameter sets for the cases (i)--(vi) are summarized in Table~\ref{tab:case_def}.
\begin{table}
\centering
\caption{
Parameter sets defining cases (i)--(vi).
}
\label{tab:case_def}
\begin{tabular}{cll}
\hline
\hline
Case & Mass-loss rate & Energy threshold \\
\hline
(i)   & conventional ($10^{-5}\,\mathrm{M_\odot\,yr^{-1}}$) & $E \ge 100\,\mathrm{TeV}$ \\
(ii)  & modern ($10^{-3}\,\mathrm{M_\odot\,yr^{-1}}$)       & $E \ge 100\,\mathrm{TeV}$ \\
(iii) & conventional & $E \ge 1\,\mathrm{TeV}$ \\
(iv)  & conventional & $E \ge 10\,\mathrm{TeV}$ \\
(v)   & modern       & $E \ge 1\,\mathrm{TeV}$ \\
(vi)  & modern       & $E \ge 10\,\mathrm{TeV}$ \\
\hline
\end{tabular}
\end{table}

\subsection{Flux above \texorpdfstring{$100\,\mathrm{TeV}$}{100 TeV}}
\begin{figure}[htbp]
    \centering
    \includegraphics[width = \linewidth, height=0.4\textheight, keepaspectratio]{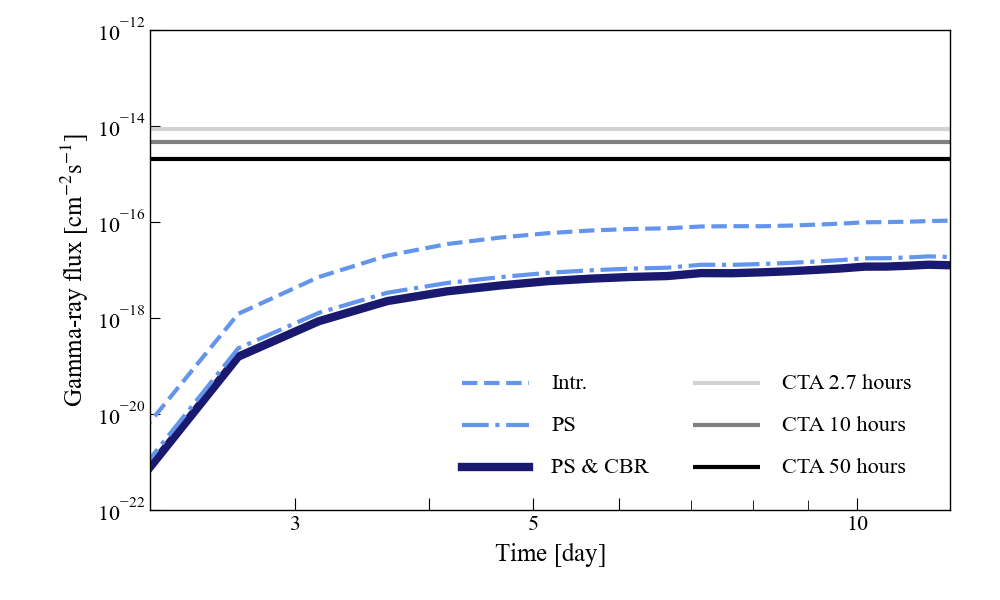}
    \caption{
    Time evolution of the integrated gamma-ray flux above $100\,{\rm TeV}$ 
    from a Type~II-P SN at $D=1\,{\rm Mpc}$ for case~(i). 
    Case~(i) adopts the conventional RSG mass-loss rate, 
    $\dot{M}_{\rm RSG}=10^{-5}\,{\rm M_\odot\,yr^{-1}}$. 
    }
    The blue dashed line shows the intrinsic flux (Intr.); 
    the dot-dashed line shows the flux attenuated by photons emitted from the SN photosphere (PS); 
    and the solid line shows the flux further attenuated by both photospheric photons and the CBR (PS+CBR). 
    The horizontal light-gray, gray, and black lines indicate the CTA sensitivities above $100\,{\rm TeV}$ 
    for $t_{\rm obs}=2.7\,{\rm h}$, $10\,{\rm h}$, and $50\,{\rm h}$, respectively.
    \label{fig:Time-flux_Ino_con_100TeV}
\end{figure}

\begin{figure}[htbp]
    \centering
    \includegraphics[width = \linewidth, height=0.4\textheight, keepaspectratio]{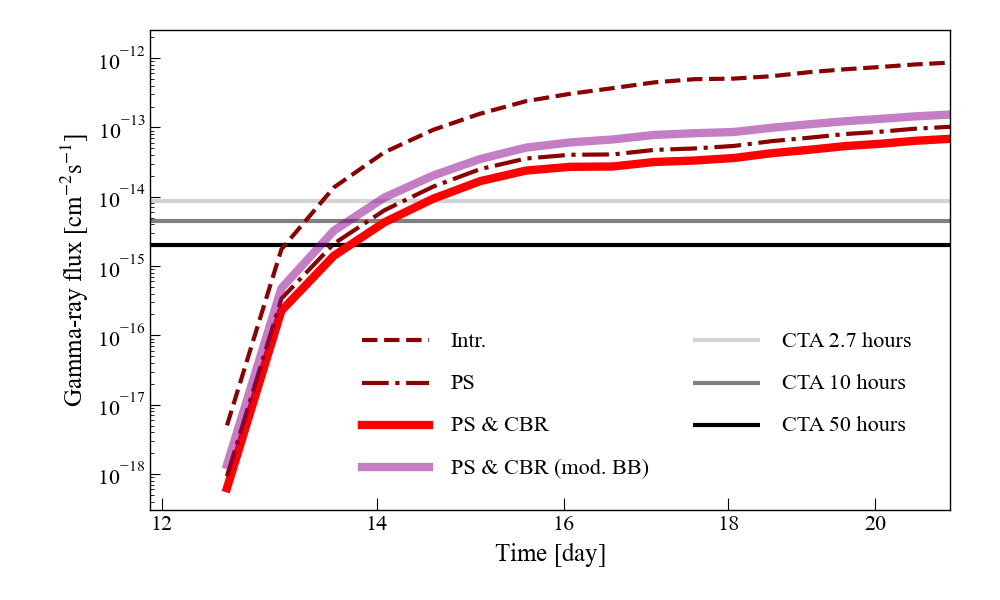}
    \caption{
    Same as Figure~\ref{fig:Time-flux_Ino_con_100TeV}, but for case~(ii). 
    Case~(ii) adopts the modern RSG mass-loss rate, 
    $\dot{M}_{\rm RSG}=10^{-3}\,{\rm M_\odot\,yr^{-1}}$, 
    and the energy threshold is $E\ge100\,{\rm TeV}$. 
    }
    The red solid line shows the flux attenuated by both photospheric photons and the CBR 
    for the fiducial blackbody photospheric model, 
    whereas the purple solid line shows the corresponding result for the diluted blackbody model.
    \label{fig:Time-flux_Ino_mod_100TeV}
\end{figure}

We first examine the gamma-ray flux above $100\,{\rm TeV}$ obtained with the kinetic-MHD simulation results \citep{2021ApJ...922....7I}, 
which constitutes the main result of this study. 
Figure~\ref{fig:Time-flux_Ino_con_100TeV} shows the flux for the conventional mass-loss rate case, $\dot{M}_{\rm RSG}=10^{-5}\,{\rm M_\odot\,yr^{-1}}$ (case i), 
and Figure~\ref{fig:Time-flux_Ino_mod_100TeV} shows the result for the modern mass-loss case, 
$\dot{M}_{\rm RSG}=10^{-3}\,{\rm M_\odot\,yr^{-1}}$ (case ii).
Throughout Section~\ref{sec:Result}, the same line-style convention is used in all figures: 
dashed for the intrinsic flux, 
dot-dashed for the flux attenuated by photospheric photons, 
and solid for the flux further attenuated by both photospheric photons and the CBR. 
From top to bottom, the horizontal lines indicate the CTA sensitivities for $t_{\rm obs}=2.7\,{\rm h}$, $10\,{\rm h}$, and $50\,{\rm h}$ observations, respectively \citep{2019scta.book.....C}. 
In cases (i) and (ii), 
the intrinsic flux rises steeply until $t\sim 5\,{\rm days}$ and $18\,{\rm days}$, respectively, and then becomes nearly constant. 
As shown in Figures~\ref{energy-f_CR_LML} and \ref{energy-f_CR}, CRs begin to reach the ${\rm PeV}$ range in the early phase, 
which drives the rapid rise of the $>100\,{\rm TeV}$ gamma-ray flux. 
After this early phase, the growth of the CR acceleration efficiency slows down and asymptotically approaches a constant value, 
leading to a corresponding flattening of the intrinsic flux 
(see Appendix~\ref{sec:E_CR_tot} for more detail). 
During the phase in which the intrinsic flux is nearly flat, 
approximately $90\%$ of the gamma-ray flux is attenuated by photospheric photons, 
with an additional $\sim 30\%$ attenuation by the CBR. 
In case (i), the attenuated flux lies below the CTA $50\,{\rm h}$ sensitivity. 
In contrast, in case (ii), which accounts for enhanced mass loss from the RSG immediately prior to the explosion, 
the flux exceeds even the $2.7\,{\rm h}$ sensitivity despite attenuation by soft photon fields. 
Since real SN photospheric spectra can be better approximated by a diluted blackbody than by a pure blackbody \citep{2014ApJ...787..157P}, 
we also tested this effect for case (ii) that is shown as a purple solid line in Figure \ref{fig:Time-flux_Ino_mod_100TeV}. 
In this calculation, the effective photospheric intensity is reduced by a factor of $0.25$ and the color temperature is increased by a factor of $1.3$ relative to the fiducial blackbody model. 
The attenuated $>100\,{\rm TeV}$ flux increases by a factor of about two, 
but the conclusion that case (ii) remains detectable with CTA is unchanged.

\subsection{Comparison with the scaling-law model}
To examine the consistency of our intrinsic-flux estimate, 
we compare it with the scaling-law model given by Equation~(10) of \citet{2020MNRAS.494.2760C}. 
A similar scaling-law estimate for the intrinsic flux from Type II-P SNe was also used by \citet{2022MNRAS.511.3321C}. 
The scaling relation adopted for this comparison is 
\begin{align}
    &F_{\gamma, {\rm intr}, {\rm C}20}(> E) 
    \approx
    6.31 \times 10^{-13} \,{\rm cm^{-2}\, s^{-1}}\,  
    \notag \\
    &\times
    \left( \frac{\phi_{0}}{0.08} \right)
    \left( \frac{\dot{M}_{\rm RSG}}{10^{-5}\, {\rm M_{\odot}\, yr^{-1}}} \right)^{2}
    \left( \frac{v_{\rm wind}}{10\, {\rm km\, s^{-1}}} \right)^{-2} 
    \notag \\
    &\times
    \left( \frac{v_{\rm Sh, 0}}{1.4 \times 10^{4}\, {\rm km s^{-1}}} \right)
    \left( \frac{D}{1\, {\rm Mpc}} \right)^{-2}
    \left( \frac{E}{1\, {\rm TeV}} \right)^{-1}
    \left( \frac{t}{1\, {\rm day}} \right)^{-1}.
    \label{eq:Intr_flux_Anal_2}
\end{align}
Here, $\phi_{0}$ and $v_{\rm Sh,0}$ denote the CR acceleration efficiency and the shock 
velocity at $t=1$~day after the explosion. 
Their model assumes that $\phi_{0}$ remains approximately constant and that the maximum CR energy exceeds $10\,{\rm PeV}$ as early as $t\sim2$--$3$~days. 
Consequently, compared with our intrinsic flux in Figure~\ref{fig:Time-flux_Ino_con_100TeV}, 
Equation~\eqref{eq:Intr_flux_Anal_2} 
tends to overestimate the flux above $100\,{\rm TeV}$ in the early phase.
Once the maximum CR energy in our simulation exceeds the ${\rm PeV}$ range, 
our model becomes consistent with Equation~\eqref{eq:Intr_flux_Anal_2}. 
For instance, in case (ii) at $t=25\,{\rm days}$, our simulation gives 
$F_{\gamma,{\rm intr}} = 1.25\times10^{-12}\,{\rm cm^{-2}\,s^{-1}}$, 
whereas Equation~\eqref{eq:Intr_flux_Anal_2} yields 
$F_{\gamma,{\rm intr},{\rm C}20} = 2.53\times10^{-12}\,{\rm cm^{-2}\,s^{-1}}$.
Thus, by incorporating the time evolution of the CR distribution function, 
our model extends the prediction of \citet{2020MNRAS.494.2760C} 
and provides a refined description of the flux evolution in the early phase.

\subsection{Fluxes above \texorpdfstring{$1$ and $10\,\mathrm{TeV}$}{1 and 10 TeV}}
\begin{figure}[htbp]
    \centering
    \includegraphics[width = \linewidth, height=0.4\textheight, keepaspectratio]{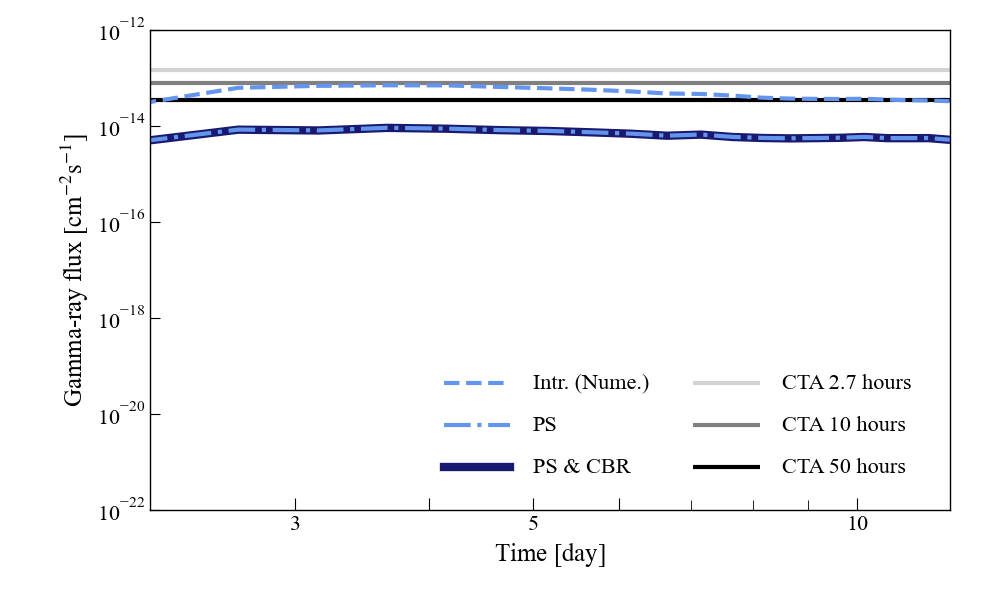}
    \caption{
    Same as Figure~\ref{fig:Time-flux_Ino_con_100TeV}, 
    but for $E\ge1\,{\rm TeV}$ with the conventional RSG mass-loss rate, 
    $\dot{M}_{\rm RSG}=10^{-5}\,{\rm M_\odot\,yr^{-1}}$ 
    (case iii).}
    \label{fig:Time-flux_Ino_con_001TeV}
\end{figure}

\begin{figure}[htbp]
    \centering
    \includegraphics[width = \linewidth, height=0.4\textheight, keepaspectratio]{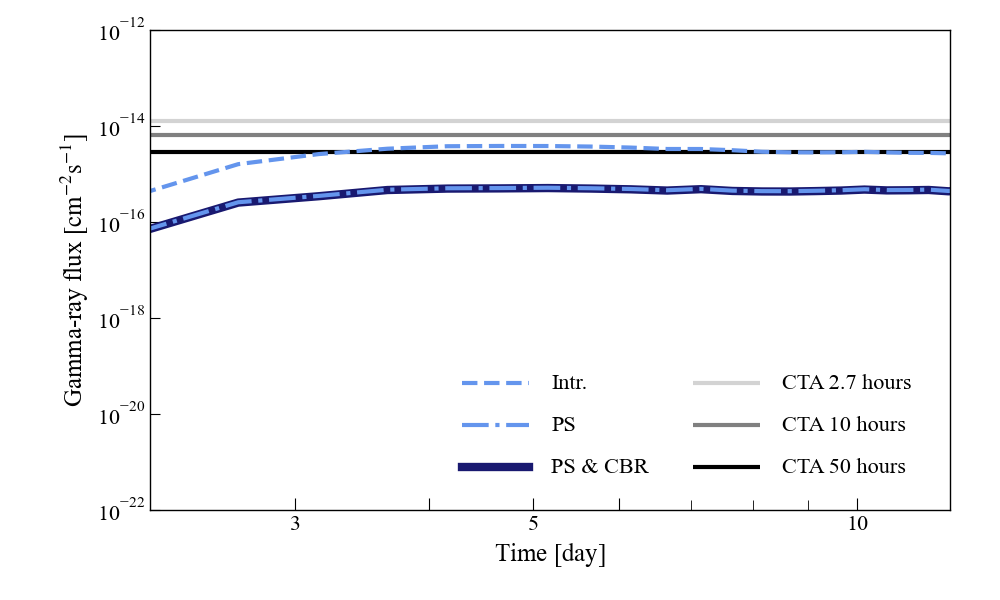}
    \caption{
    Same as Figure~\ref{fig:Time-flux_Ino_con_100TeV}, 
    but for $E\ge10\,{\rm TeV}$ with the conventional RSG mass-loss rate, 
    $\dot{M}_{\rm RSG}=10^{-5}\,{\rm M_\odot\,yr^{-1}}$ 
    (case iv).}
    \label{fig:Time-flux_Ino_con_010TeV}
\end{figure}

\begin{figure}[htbp]
    \centering
    \includegraphics[width = \linewidth, height=0.4\textheight, keepaspectratio]{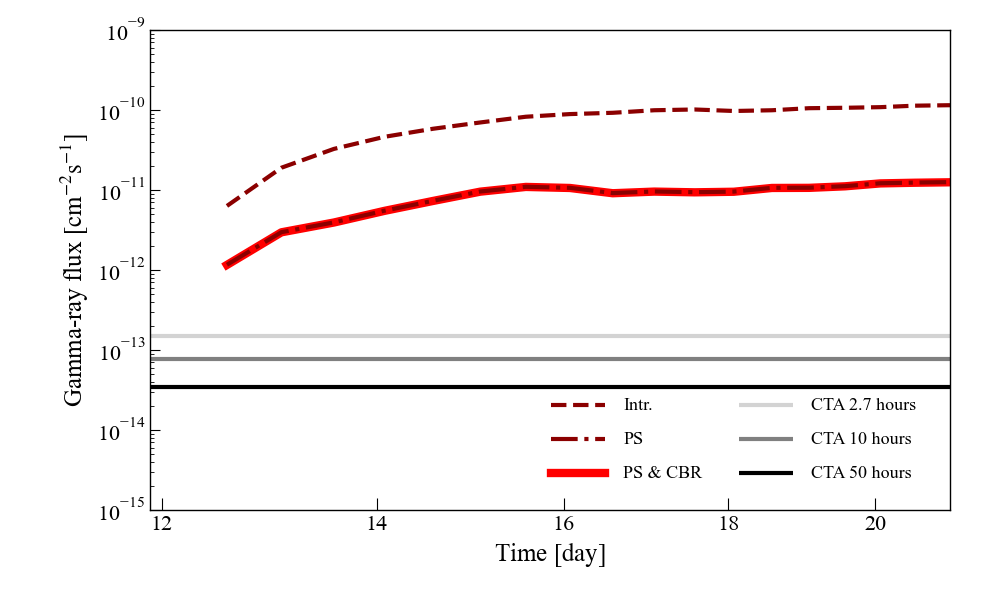}
    \caption{
    Same as Figure~\ref{fig:Time-flux_Ino_mod_100TeV}, 
    but for $E\ge1\,{\rm TeV}$ with the modern RSG mass-loss rate, 
    $\dot{M}_{\rm RSG}=10^{-3}\,{\rm M_\odot\,yr^{-1}}$ 
    (case~v).
    }
    \label{fig:Time-flux_Ino_mod_001TeV}
\end{figure}

\begin{figure}[htbp]
    \centering
    \includegraphics[width = \linewidth, height=0.4\textheight, keepaspectratio]{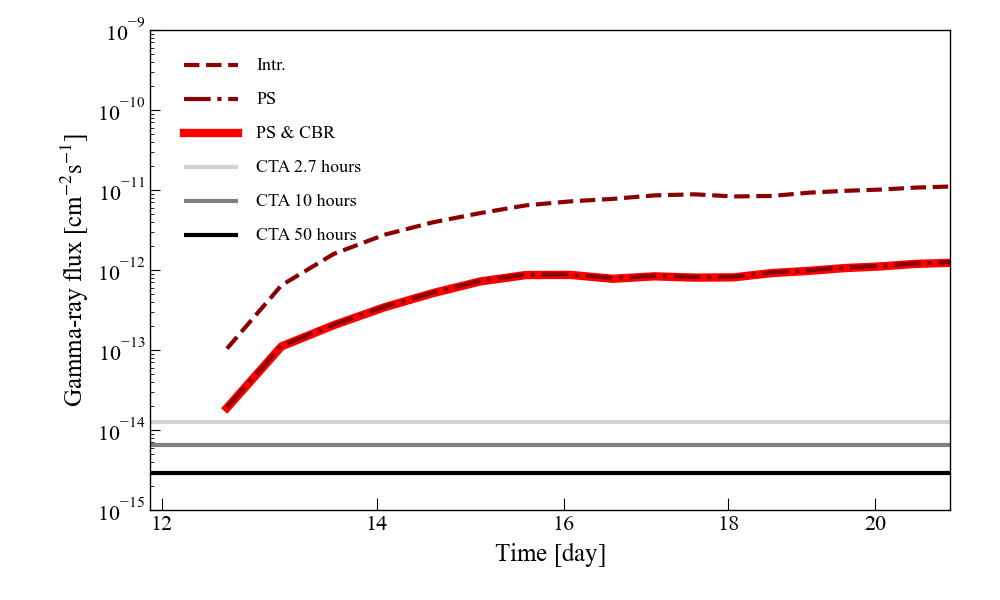}
    \caption{
    Same as Figure~\ref{fig:Time-flux_Ino_mod_100TeV}, 
    but for $E\ge10\,{\rm TeV}$ with the modern RSG mass-loss rate, 
    $\dot{M}_{\rm RSG}=10^{-3}\,{\rm M_\odot\,yr^{-1}}$ 
    (case~vi).}
    \label{fig:Time-flux_Ino_mod_010TeV}
\end{figure}

We next examine the fluxes at lower energy thresholds, $E\ge 1\,{\rm TeV}$ and $E\ge 10\,{\rm TeV}$, computed with the kinetic-MHD simulation model.
Figures~\ref{fig:Time-flux_Ino_con_001TeV} and ~\ref{fig:Time-flux_Ino_con_010TeV} 
present the fluxes for $E\ge 1\,{\rm TeV}$ (case iii) and $E\ge 10\,{\rm TeV}$ (case iv), respectively, 
assuming the conventional mass-loss rate. 
Similarly, Figures~\ref{fig:Time-flux_Ino_mod_001TeV} and ~\ref{fig:Time-flux_Ino_mod_010TeV} 
show the fluxes for $E\ge 1\,{\rm TeV}$ (case v) and $E\ge 10\,{\rm TeV}$ (case vi) under the modern mass-loss rate. 
In all four panels, the horizontal lines indicate the CTA sensitivities for the respective energy thresholds: 
light gray, gray, and black correspond to $t_{\rm obs}=2.7\,{\rm h}$, $10\,{\rm h}$, and $50\,{\rm h}$ observations, respectively.

At $t = 10\,{\rm days}$ with the conventional mass-loss rate, the intrinsic flux 
for $E \ge 1\,{\rm TeV}$ (case iii) and $E \ge 10\,{\rm TeV}$ (case iv) 
exceeds that for $E \ge 100\,{\rm TeV}$ (case i) 
by approximately two orders and one order of magnitude, respectively. 
A similar trend appears at $t = 20\,{\rm days}$ with the modern mass-loss rate for cases (v) and (vi). 
In all four lower-energy cases (iii--vi), attenuation by photospheric photons reduces the flux 
by about one order of magnitude, while attenuation by the CBR is negligible. 
For the modern mass-loss rate, the attenuated flux remains above the CTA $t_{\rm obs}=2.7\,{\rm h}$ sensitivity, 
and above the $E\ge 1\,{\rm TeV}$ threshold it is detectable from very early times. 

\section{Discussion}\label{Discussion}
In this section, we evaluate the maximum detectable distance, $D_{\rm lim}(E,t_{\rm obs})$, 
for CTA observations of gamma-rays from Type II-P SNe under a given observation time $t_{\rm obs}$. 
We also estimate the corresponding event rates by combining 
$D_{\rm lim}$ with the core-collapse SN rates of nearby galaxies. 
However, these event-rate estimates should be regarded as conditional estimates 
based on the adopted CSM assumptions, rather than as general predictions for 
the entire Type II-P SN population.

CTA provides the deepest sensitivity in the $\sim1$--$100\,\mathrm{TeV}$ energy range 
for observation times of several tens of hours, making it well suited for searching 
for gamma-rays produced by ${\rm PeV}$ CRs in very young supernova remnants. 
One might expect that LHAASO would be a better instrument for detecting 
$100\,\mathrm{TeV}$ gamma-rays. 
However, for transient objects with a $50\,{\rm h}$ integration time, 
CTA achieves roughly $20$ times higher sensitivity than LHAASO 
\citep{2019scta.book.....C, 2025arXiv251216638T}. 
We consider three energy thresholds, 
$E \ge 1\,\mathrm{TeV}$, $E \ge 10\,\mathrm{TeV}$, and $E \ge 100\,\mathrm{TeV}$.

In this work, we focus on gamma-rays in the $1$--$100\,{\rm TeV}$ range, especially on $100\,{\rm TeV}$ gamma-rays, because they are closely related to the presence of CR protons accelerated to PeV energies in the very early phase of SNe.
A broader study covering GeV to PeV gamma-rays would be useful for constraining the overall CR spectrum and its time evolution.
In addition, observations with other gamma-ray instruments, such as Fermi-LAT, H.E.S.S., HAWC, LHAASO, and CTA, would provide complementary information over different energy ranges and observational timescales.
Such a broad-band, multi-instrument study is beyond the scope of the present work and is left for future work.

Hadronic interactions also produce neutrinos through charged-pion decay.
Such neutrino emission would provide an important multi-messenger probe of PeV CR acceleration in very young SNe.
However, evaluating the neutrino detectability with IceCube requires calculations of the neutrino spectrum and flavor composition, including the effect of neutrino oscillations during propagation to Earth, and comparison with the energy- and flavor-dependent IceCube sensitivity.
Therefore, a detailed study of the neutrino flux and IceCube detectability is reserved for a separate paper.

Our procedure is as follows: 
(1) we define the maximum detectable distance $D_{\rm lim}(E,t_{\rm obs})$ 
as a function of the gamma-ray energy threshold $E$ and observation time $t_{\rm obs}$; 
(2) we then estimate the corresponding event rates by summing the core-collapse SN rates 
of galaxies within $D_{\rm lim}$.

\begin{figure}[htbp]
    \centering
    \includegraphics[width = \linewidth, height=0.6\textheight, keepaspectratio]{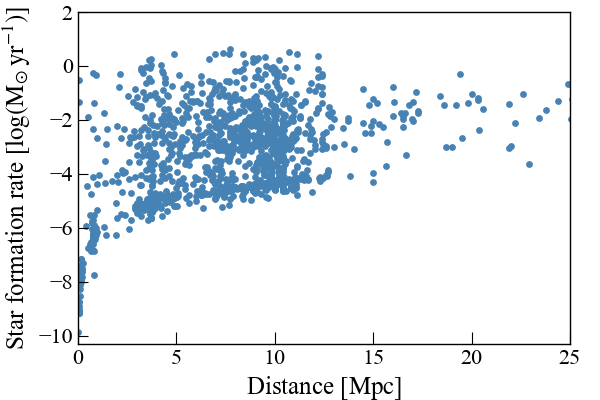}
    \caption{
    Distances and star formation rates of nearby galaxies within $25\,{\rm Mpc}$,
    taken from the galaxy catalog of 
    \citet{2013AJ....145..101K}.
    }
    \label{fig:D-SFR_Karachentsev+13}
\end{figure}

We define $D_{\rm lim}(E,t_{\rm obs})$ as the distance at which the integrated gamma-ray flux above $E$, after attenuation by photospheric photons and the CBR, equals the CTA sensitivity for $t_{\rm obs}$. 
Since the CTA sensitivities differ between the southern and northern sites, 
we evaluate the detectability for both sites. 
We compute the detection rates separately for the two sites by accounting for their respective sky coverage, and then sum them to obtain the all-sky detection rates. 
The detectable distances estimated in this work are based on the published CTA sensitivities for the adopted energy thresholds and observation times.
In actual transient observations, the effective sensitivity can depend on the zenith angle, background conditions, source visibility, and achievable exposure time after the SN discovery.
Therefore, the values of $D_{\rm lim}$ and the corresponding event-rate estimates should be regarded as fiducial estimates based on the adopted CTA sensitivities and observing conditions.
We estimate the core-collapse SN rates in nearby galaxies 
assuming the Chabrier initial mass function (IMF) \citep{2003PASP..115..763C}. 
From the IMF, $f_M=17.6\%$ of the total stellar mass is converted into stars with masses between $8\,{\rm M_\odot}$ and $50\,{\rm M_\odot}$, with an average mass $\bar{M}_{\rm SN}\approx16.2\,{\rm M_\odot}$. 
Therefore, the core-collapse SN rate in a galaxy with a star formation rate ${\rm SFR}$ is estimated as ${\rm SFR}\times f_M/\bar{M}_{\rm SN}$. 
We adopt galaxy distances and star formation rates from the catalog of 
\citet{2013AJ....145..101K}, 
and show their distribution in Figure~\ref{fig:D-SFR_Karachentsev+13}. 

This procedure also introduces uncertainties in the event-rate estimates.
In addition, the event rates relevant to the dense CSM cases depend on the fraction of core-collapse SNe that are Type~II-P SNe and on the occurrence fraction of Type~II-P SNe with sufficiently dense and extended CSM.
These fractions are not well constrained at present.
Thus, the event-rate estimates presented in this work should be interpreted as conditional and fiducial estimates rather than robust predictions for the entire Type~II-P SN population.

\begin{figure}[htbp]
    \centering
    \includegraphics[width = \linewidth, height=0.6\textheight, keepaspectratio]{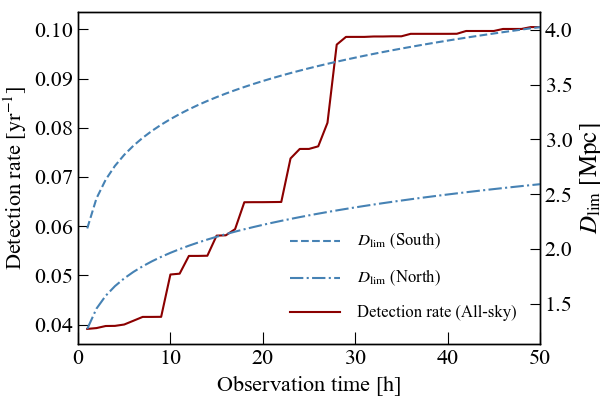}
    \caption{
    Maximum detectable distance, $D_{\rm lim}$, for the CTA southern and northern sites, 
    and the all-sky gamma-ray detection rate as a function of the observation time for case (ii) 
    ($\dot{M}_{\rm RSG}=10^{-3}\,\mathrm{M_\odot\,yr^{-1}}$; $E \ge 100\,\mathrm{TeV}$).  
    The dashed and dot-dashed lines show $D_{\rm lim}$ for the southern and northern sites, respectively, and the solid line shows the all-sky detection rate. 
    }
    \label{fig:t-obs-DR}
\end{figure}

Figure~\ref{fig:t-obs-DR} shows $D_{\rm lim}$ and the corresponding gamma-ray detection rate for case~(ii) as a function of the observation time $t_{\rm obs}$. 
Tables~\ref{tab:detection_case_CMLR_100TeV}--\ref{tab:detection_case_MMLR_010TeV} summarize $D_{\rm lim}$ and the detection rates as a function of $t_{\rm obs}$ for all cases (i)--(vi), including separate estimates for the southern and northern CTA sites as well as the combined all-sky rate. 

In particular, for case (ii), 
a $50\,{\rm h}$ CTA observation gives $D_{\rm lim}\sim 4\,{\rm Mpc}$ for $E\ge100\,{\rm TeV}$. 
This result indicates that, if a nearby Type II-P SN satisfying the dense and extended CSM conditions assumed in Model $1$ occurs, 
CTA can detect its $100\,{\rm TeV}$ gamma-ray emission out to a few Mpc. 
The corresponding all-sky event-rate estimate is one event per $\sim 10$ years.

The importance of the high-density CSM is also evident from comparison with 
the conventional mass-loss case at the same energy threshold. 
For $E\ge100\,{\rm TeV}$, case (i) yields only 
$D_{\rm lim}\sim 0.1\,{\rm Mpc}$ for a $50\,{\rm h}$ observation, 
and its all-sky detection rate remains at the level of one event per $\sim 40$ years. 
Thus, the enhanced detectability in case (ii) mainly reflects the increase in the 
gamma-ray luminosity caused by the high-density CSM.

The CSM density also depends on the wind velocity adopted in the RSG wind model.
As shown in Equation~\ref{eq:rho_CSM}, for a fixed mass-loss rate, a larger $v_{\rm wind}$ gives a lower CSM density and hence weaker hadronic gamma-ray production, whereas a smaller $v_{\rm wind}$ gives a higher CSM density and stronger gamma-ray production. 
In this work, we adopt $v_{\rm wind}=10\,{\rm km\,s^{-1}}$ as a typical value for an RSG wind. 
However, in self-consistent kinetic-MHD simulations, changing $v_{\rm wind}$ would affect not only the hadronic target density but also CR acceleration, magnetic-field amplification, and free--free cooling through the change in the CSM density. 
Therefore, the dependence of the gamma-ray flux and detectability 
on $v_{\rm wind}$ cannot be fully evaluated by a simple density scaling alone. 
A quantitative evaluation of the wind-velocity dependence requires kinetic-MHD simulations 
with different values of $v_{\rm wind}$, and is left for future work.

The intrinsic gamma-ray flux does not necessarily increase monotonically with the mass-loss rate.
In the conventional mass-loss case, the CSM density is low and hadronic gamma-ray production is inefficient, resulting in a small detectable distance for $100\,{\rm TeV}$ gamma-rays.
In the modern mass-loss case adopted in Model $1$, the higher CSM density increases the hadronic target density and greatly improves the gamma-ray detectability.
However, if the mass-loss rate, or equivalently the CSM density, becomes even larger, free--free cooling can become important and suppress PeV CR acceleration, as discussed in Section~\ref{sec:CR&CSM}. 
In such a case, the $100\,{\rm TeV}$ gamma-ray flux can decrease despite the larger target density. 
Therefore, the dense CSM cases in this work should be interpreted as conditional cases in which the CSM is sufficiently dense to enhance hadronic gamma-ray production, but not so dense that free--free cooling strongly suppresses PeV CR acceleration. 
The occurrence fraction of Type~II-P SNe satisfying such conditions is still uncertain, and should be constrained by future early-time SN observations and additional kinetic-MHD simulations with different CSM densities.

The results at lower energy thresholds further show the same trend. 
The conventional mass-loss cases at these thresholds, cases (iii) and (iv), 
yield relatively small detectable distances, 
$D_{\rm lim}\lesssim 0.4\,{\rm Mpc}$ even for a $50\,{\rm h}$ observation, 
and their all-sky detection rates remain at the level of one event per $\sim 40$ years. 
By contrast, the high-density CSM cases, cases (v) and (vi), give much larger detectable distances for 
$E\ge1\,{\rm TeV}$ and $E\ge10\,{\rm TeV}$, reaching 
$D_{\rm lim}\sim 15$--$23\,{\rm Mpc}$ for a $50\,{\rm h}$ observation. 
The corresponding all-sky detection rates are close to one event per year.

 These detectable distances in the high-density CSM cases are larger than those inferred by 
\citet{2022MNRAS.511.3321C} for Type~II-P RSG winds. 
They investigated the detectability of early gamma-ray emission from Type~II-P SNe 
for RSG wind mass-loss rates 
$\dot{M}_{\rm RSG}=10^{-8}$--$10^{-5}\,{\rm M_\odot\, yr^{-1}}$. 
For emission above $1\,{\rm TeV}$, their predicted fluxes at 
$D=1\,{\rm Mpc}$ are below the CTA sensitivity for a $50\,{\rm h}$ observation 
around $t\sim 10$ days, implying that the detectability horizon is 
$\lesssim 1\,{\rm Mpc}$ for Type~II-P RSG winds. 
The larger detectable distances obtained in our high-density CSM cases 
mainly arise from the larger intrinsic gamma-ray luminosity. 
Since the intrinsic flux 
approximately scales as $F_{\gamma,{\rm intr}}\propto \dot{M}_{\rm RSG}^{2}$ 
(see Equation~\ref{eq:Intr_flux_Anal_2}), 
increasing the mass-loss rate from $\sim 10^{-5}$ to $10^{-3}\,{\rm M_\odot\, yr^{-1}}$ substantially enhances the 
gamma-ray emissivity and increase the detectability horizon. 
However, the observed flux is still reduced by $\gamma\gamma$ absorption, and 
the degree of attenuation can differ from that in 
\citet{2022MNRAS.511.3321C} because of the different modelling of the 
photospheric emission and the shock--photosphere geometry. 
In particular, the ratio of the forward-shock radius to the photospheric 
radius is larger in our model than that in \citet{2022MNRAS.511.3321C}, 
leading to the reduced attenuation by photospheric photons.

Recent observations support the presence of dense CSM around many Type~II SN progenitors, although its radial extent and occurrence fraction are still uncertain. 
\citet{2018NatAs...2..808F} showed that the early optical light curves of many Type~II SNe are consistent with dense CSM produced by enhanced mass loss, often with $\dot{M}_{\rm RSG} \gtrsim10^{-4}\,{\rm M_\odot\, yr^{-1}}$.
The models of \citet{2018NatAs...2..808F} include enhanced mass loss extending to 
$\sim10^{15}\,{\rm cm}$, corresponding to a shock-crossing time of 
$\sim10$ days for $v_{\rm sh}=10^4\,{\rm km\,s^{-1}}$.
On the other hand, \citet{2023ApJ...952..119B} found that the spectroscopic 
signatures of confined CSM in most Type~II SNe persist for only $\sim5$ days, 
while longer-lived events with timescales $>10$ days are rarer. 
This suggests that CSM dense enough and extended enough to affect the 
emission over $\sim10$ days may correspond to a subset of Type~II SNe. 
Taken together, these observations motivate considering a dense CSM case, while also indicating that the radial extent of such dense CSM is uncertain.

To quantify the influence of the radial extent of the dense CSM, 
we additionally performed a parametric estimate for case (ii). 
In Figure \ref{fig:t-obs-DR}, we estimate $D_{\rm lim}$ at $t = 25\, {\rm days}$, 
which corresponds to the time when the forward shock reaches $r \approx 3 \times 10^{15}\, {\rm cm}$. 
If we estimate $D_{\rm lim}$ at an earlier time, 
we can obtain the maximum detectable distance for cases with a smaller CSM radial extent.
If we denote the radial extent of the CSM as $R_{\rm CSM}$, 
the corresponding evaluation time is calculated as $t = R_{\rm CSM}/v_{\rm Sh}$.
Assuming a $50\,{\rm h}$ observation with CTA South, 
we find that the maximum detectable distance for $E\ge100\,{\rm TeV}$ 
is changed to $D_{\rm lim}=2.89\,{\rm Mpc}$ 
for $R_{\rm CSM} = 2.2\times10^{15}\,{\rm cm}$ 
to $D_{\rm lim}=3.35\,{\rm Mpc}$ 
for $R_{\rm CSM} = 2.5\times10^{15}\,{\rm cm}$, 
and to $D_{\rm lim}=3.74\,{\rm Mpc}$ 
for $R_{\rm CSM}=2.8\times10^{15}\,{\rm cm}$.
The corresponding event-rate estimates are approximately one event per $23$, $22$, and $11$ years, respectively.
The rapid increase in the event rate between $R_{\rm CSM}=2.5\times10^{15}\,{\rm cm}$ and $2.8\times10^{15}\,{\rm cm}$ occurs because the detectable distance reaches the distance range of nearby galaxies with high star formation rates, such as M82, NGC 253, and M81.
These results show that an extended dense CSM 
is important for the detectability of $100\,{\rm TeV}$ gamma-ray emission, 
but that the dense CSM does not necessarily have to 
extend to $R_{\rm CSM} = 3\times10^{15}\,{\rm cm}$ to produce a detectable flux.

In addition to the uncertainties in the CSM properties, the predicted intrinsic gamma-ray luminosity also depends on the adopted CR distribution function. 
The CR distribution function used in this work is taken from kinetic-MHD simulation results obtained with a fixed injection fraction, $\eta_{\rm inj}=6\times10^{-4}$. 
Therefore, the dependence on $\eta_{\rm inj}$ remains as an uncertainty in the present calculation.
Our flux calculation includes CRs distributed around the shock front, including those propagating into the upstream region contained in the radial profile of $f(r,p)$. 
A quantitative study of the $\eta_{\rm inj}$ dependence is left for future work.

\begin{table*}
\centering
\small
\caption{
Maximum detectable distance and corresponding gamma-ray detection rate as a function of CTA observation time for case (i) ($\dot{M}_{\rm RSG}=10^{-5}\,\mathrm{M_\odot\,yr^{-1}}$; $E \ge 100\,\mathrm{TeV}$).
}
\label{tab:detection_case_CMLR_100TeV}
\begin{tabular}{ccccccc}\hline\hline
CTA obs. time & $D_{\rm lim}$ (South) & $D_{\rm lim}$ (North) & Detection rate & Detection rate & Detection rate & Mean interval \\
$[\mathrm{h}]$ & $[\mathrm{Mpc}]$ & $[\mathrm{Mpc}]$ & (South) $[\mathrm{yr^{-1}}]$ & (North) $[\mathrm{yr^{-1}}]$ & (All-sky) $[\mathrm{yr^{-1}}]$ & (All-sky) $[\mathrm{yr/event}]$ \\
\hline
3 & 0.0523 & 0.0254 & $1.33 \times 10^{-2}$ & $1.00 \times 10^{-2}$ & $2.33 \times 10^{-2}$ & 42.9 \\
5 & 0.0594 & 0.0289 & $1.33 \times 10^{-2}$ & $1.00 \times 10^{-2}$ & $2.33 \times 10^{-2}$ & 42.9 \\
10 & 0.0704 & 0.0343 & $1.38 \times 10^{-2}$ & $1.00 \times 10^{-2}$ & $2.38 \times 10^{-2}$ & 42.0 \\
30 & 0.0923 & 0.0450 & $1.38 \times 10^{-2}$ & $1.00 \times 10^{-2}$ & $2.38 \times 10^{-2}$ & 42.0 \\
50 & 0.105 & 0.0511 & $1.38 \times 10^{-2}$ & $1.00 \times 10^{-2}$ & $2.38 \times 10^{-2}$ & 42.0 \\
\hline
\end{tabular}
\end{table*}
\begin{table*}
\centering
\small
\caption{
Same as Table~\ref{tab:detection_case_CMLR_100TeV} but for case (ii) ($\dot{M}_{\rm RSG}=10^{-3}\,\mathrm{M_\odot\,yr^{-1}}$; $E \ge 100\,\mathrm{TeV}$).
}
\label{tab:detection_case_MMLR_100TeV}
\begin{tabular}{ccccccc}\hline\hline
CTA obs. time & $D_{\rm lim}$ (South) & $D_{\rm lim}$ (North) & Detection rate & Detection rate & Detection rate & Mean interval \\
$[\mathrm{h}]$ & $[\mathrm{Mpc}]$ & $[\mathrm{Mpc}]$ & (South) $[\mathrm{yr^{-1}}]$ & (North) $[\mathrm{yr^{-1}}]$ & (All-sky) $[\mathrm{yr^{-1}}]$ & (All-sky) $[\mathrm{yr/event}]$ \\
\hline
3 & 2.63 & 1.57 & $2.76 \times 10^{-2}$ & $1.63 \times 10^{-2}$ & $4.39 \times 10^{-2}$ & 22.8 \\
5 & 2.86 & 1.73 & $2.81 \times 10^{-2}$ & $1.63 \times 10^{-2}$ & $4.44 \times 10^{-2}$ & 22.5 \\
10 & 3.18 & 1.96 & $2.85 \times 10^{-2}$ & $1.63 \times 10^{-2}$ & $4.48 \times 10^{-2}$ & 22.3 \\
30 & 3.75 & 2.38 & $7.78 \times 10^{-2}$ & $1.63 \times 10^{-2}$ & $9.41 \times 10^{-2}$ & 10.6 \\
50 & 4.03 & 2.59 & $7.87 \times 10^{-2}$ & $1.63 \times 10^{-2}$ & $9.50 \times 10^{-2}$ & 10.5 \\
\hline
\end{tabular}
\end{table*}
\begin{table*}
\centering
\small
\caption{
Same as Table~\ref{tab:detection_case_CMLR_100TeV} but for case (iii) ($\dot{M}_{\rm RSG}=10^{-5}\,\mathrm{M_\odot\,yr^{-1}}$; $E \ge 1\,\mathrm{TeV}$).
}
\label{tab:detection_case_CMLR_001TeV}
\begin{tabular}{ccccccc}\hline\hline
CTA obs. time & $D_{\rm lim}$ (South) & $D_{\rm lim}$ (North) & Detection rate & Detection rate & Detection rate & Mean interval \\
$[\mathrm{h}]$ & $[\mathrm{Mpc}]$ & $[\mathrm{Mpc}]$ & (South) $[\mathrm{yr^{-1}}]$ & (North) $[\mathrm{yr^{-1}}]$ & (All-sky) $[\mathrm{yr^{-1}}]$ & (All-sky) $[\mathrm{yr/event}]$ \\
\hline
3 & 0.175 & 0.143 & $1.38 \times 10^{-2}$ & $1.00 \times 10^{-2}$ & $2.38 \times 10^{-2}$ & 42.0 \\
5 & 0.199 & 0.163 & $1.38 \times 10^{-2}$ & $1.00 \times 10^{-2}$ & $2.38 \times 10^{-2}$ & 42.0 \\
10 & 0.237 & 0.193 & $1.38 \times 10^{-2}$ & $1.00 \times 10^{-2}$ & $2.38 \times 10^{-2}$ & 42.0 \\
30 & 0.312 & 0.254 & $1.38 \times 10^{-2}$ & $1.00 \times 10^{-2}$ & $2.38 \times 10^{-2}$ & 42.0 \\
50 & 0.354 & 0.289 & $1.38 \times 10^{-2}$ & $1.00 \times 10^{-2}$ & $2.38 \times 10^{-2}$ & 42.0 \\
\hline
\end{tabular}
\end{table*}
\begin{table*}
\centering
\small
\caption{
Same as Table~\ref{tab:detection_case_CMLR_100TeV} but for case (iv) ($\dot{M}_{\rm RSG}=10^{-5}\,\mathrm{M_\odot\,yr^{-1}}$; $E \ge 10\,\mathrm{TeV}$).
}
\label{tab:detection_case_CMLR_010TeV}
\begin{tabular}{ccccccc}\hline\hline
CTA obs. time & $D_{\rm lim}$ (South) & $D_{\rm lim}$ (North) & Detection rate & Detection rate & Detection rate & Mean interval \\
$[\mathrm{h}]$ & $[\mathrm{Mpc}]$ & $[\mathrm{Mpc}]$ & (South) $[\mathrm{yr^{-1}}]$ & (North) $[\mathrm{yr^{-1}}]$ & (All-sky) $[\mathrm{yr^{-1}}]$ & (All-sky) $[\mathrm{yr/event}]$ \\
\hline
3 & 0.187 & 0.116 & $1.38 \times 10^{-2}$ & $1.00 \times 10^{-2}$ & $2.38 \times 10^{-2}$ & 42.0 \\
5 & 0.213 & 0.132 & $1.38 \times 10^{-2}$ & $1.00 \times 10^{-2}$ & $2.38 \times 10^{-2}$ & 42.0 \\
10 & 0.253 & 0.157 & $1.38 \times 10^{-2}$ & $1.00 \times 10^{-2}$ & $2.38 \times 10^{-2}$ & 42.0 \\
30 & 0.333 & 0.207 & $1.38 \times 10^{-2}$ & $1.00 \times 10^{-2}$ & $2.38 \times 10^{-2}$ & 42.0 \\
50 & 0.378 & 0.235 & $1.38 \times 10^{-2}$ & $1.00 \times 10^{-2}$ & $2.38 \times 10^{-2}$ & 42.0 \\
\hline
\end{tabular}
\end{table*}
\begin{table*}
\centering
\small
\caption{
Same as Table~\ref{tab:detection_case_CMLR_100TeV} but for case (v) ($\dot{M}_{\rm RSG}=10^{-3}\,\mathrm{M_\odot\,yr^{-1}}$; $E \ge 1\,\mathrm{TeV}$).
}
\label{tab:detection_case_MMLR_001TeV}
\begin{tabular}{ccccccc}\hline\hline
CTA obs. time & $D_{\rm lim}$ (South) & $D_{\rm lim}$ (North) & Detection rate & Detection rate & Detection rate & Mean interval \\
$[\mathrm{h}]$ & $[\mathrm{Mpc}]$ & $[\mathrm{Mpc}]$ & (South) $[\mathrm{yr^{-1}}]$ & (North) $[\mathrm{yr^{-1}}]$ & (All-sky) $[\mathrm{yr^{-1}}]$ & (All-sky) $[\mathrm{yr/event}]$ \\
\hline
3 & 9.63 & 7.97 & $36.7 \times 10^{-2}$ & $23.6 \times 10^{-2}$ & $60.3 \times 10^{-2}$ & 1.66 \\
5 & 10.8 & 8.98 & $51.2 \times 10^{-2}$ & $29.7 \times 10^{-2}$ & $80.9 \times 10^{-2}$ & 1.24 \\
10 & 12.7 & 10.5 & $58.2 \times 10^{-2}$ & $35.0 \times 10^{-2}$ & $93.2 \times 10^{-2}$ & 1.07 \\
30 & 16.2 & 13.6 & $58.5 \times 10^{-2}$ & $38.7 \times 10^{-2}$ & $97.2 \times 10^{-2}$ & 1.03 \\
50 & 18.2 & 15.2 & $58.8 \times 10^{-2}$ & $38.9 \times 10^{-2}$ & $97.7 \times 10^{-2}$ & 1.02 \\
\hline
\end{tabular}
\end{table*}
\begin{table*}
\centering
\small
\caption{
Same as Table~\ref{tab:detection_case_CMLR_100TeV} but for case (vi) ($\dot{M}_{\rm RSG}=10^{-3}\,\mathrm{M_\odot\,yr^{-1}}$; $E \ge 10\,\mathrm{TeV}$).
}
\label{tab:detection_case_MMLR_010TeV}
\begin{tabular}{ccccccc}\hline\hline
CTA obs. time & $D_{\rm lim}$ (South) & $D_{\rm lim}$ (North) & Detection rate & Detection rate & Detection rate & Mean interval \\
$[\mathrm{h}]$ & $[\mathrm{Mpc}]$ & $[\mathrm{Mpc}]$ & (South) $[\mathrm{yr^{-1}}]$ & (North) $[\mathrm{yr^{-1}}]$ & (All-sky) $[\mathrm{yr^{-1}}]$ & (All-sky) $[\mathrm{yr/event}]$ \\
\hline
3 & 11.6 & 7.29 & $54.4 \times 10^{-2}$ & $14.7 \times 10^{-2}$ & $69.1 \times 10^{-2}$ & 1.45 \\
5 & 13.2 & 8.26 & $58.2 \times 10^{-2}$ & $23.7 \times 10^{-2}$ & $81.9 \times 10^{-2}$ & 1.22 \\
10 & 15.6 & 9.79 & $58.4 \times 10^{-2}$ & $32.8 \times 10^{-2}$ & $91.2 \times 10^{-2}$ & 1.10 \\
30 & 20.3 & 12.8 & $59.6 \times 10^{-2}$ & $38.7 \times 10^{-2}$ & $98.3 \times 10^{-2}$ & 1.02 \\
50 & 23.0 & 14.5 & $59.7 \times 10^{-2}$ & $38.8 \times 10^{-2}$ & $98.5 \times 10^{-2}$ & 1.02 \\
\hline
\end{tabular}
\end{table*}

\section{Summary}\label{Summary}
We have evaluated the gamma-ray flux in the very early phase of CCSNe. 
Our analysis is based on the early-time CR acceleration process computed by the kinetic-MHD simulations of \citet{2021ApJ...922....7I}, incorporating gamma-ray attenuation due to interactions with both photospheric and CBR photons. 
For Type~II-P SNe, we found that the intrinsic gamma-ray flux in the $1$--$100\,{\rm TeV}$ range is attenuated by approximately one order of magnitude due to interactions with photospheric photons. 
Furthermore, for a source located at a distance of $1\,{\rm Mpc}$, the $100\,{\rm TeV}$ gamma-ray flux is additionally reduced by about two-thirds due to interactions with CBR photons. 
For sources located beyond $\sim 1\,{\rm Mpc}$, attenuation by the CBR becomes substantial, 
especially at $100\,{\rm TeV}$, and should be included in flux predictions.

Previous work by \citet{2020MNRAS.494.2760C} suggested that, even with a $50\,{\rm h}$ observation by CTA, the possibility of detecting gamma-ray emission from SN~1993J within approximately $20\,{\rm days}$ after core-collapse is significantly suppressed when $\gamma\gamma$ absorption is considered. 
However, as shown in Figures~\ref{fig:Time-flux_Ino_mod_100TeV}, 
\ref{fig:Time-flux_Ino_mod_001TeV}, and~\ref{fig:Time-flux_Ino_mod_010TeV}, 
assuming a dense CSM produced by an enhanced RSG mass-loss rate of $\sim 10^{-3}\,{\rm M_{\odot}\,yr^{-1}}$, the expected gamma-ray flux becomes substantially higher than in the conventional mass-loss case. 
Our results suggest that observing the brightening of the gamma-ray flux within the first $\sim 20\,{\rm days}$ after core-collapse provides a valuable probe of CR acceleration up to ${\rm PeV}$ energies.

For the dense CSM case, we find that CTA can detect $100\,{\rm TeV}$ gamma-rays with a $50\,{\rm h}$ observation out to $D_{\rm lim}\sim4.0\,{\rm Mpc}$. 
This detectable distance applies to a Type~II-P SN satisfying the dense and extended CSM conditions adopted in this work, and should not be interpreted as a general prediction for all Type~II-P SNe. 
The corresponding event-rate estimates are therefore conditional on the adopted CSM assumptions, as discussed in Section~\ref{Discussion}.
As discussed in Section~\ref{sec:CR&CSM}, if the CSM density is substantially higher than assumed in our model, free--free cooling can suppress PeV CR acceleration and reduce the $100\,{\rm TeV}$ gamma-ray flux. 
Future early-time observations and modeling of Type~II SNe will be important for constraining the density structure, radial extent, and occurrence fraction of such dense CSM environments.

\begin{acknowledgments}
\indent
We thank S. Inoue, S. Kimura, T. Moriya, K. Murase and K. Kashiyama for fruitful discussions that greatly advanced this work. 
This work is supported by Grants-in-Aid from the Ministry of Education, Culture, Sports, Science, and Technology (MEXT) of Japan grant Nos. 20H01944 and 23H01211. 
This work was performed using the super-computing facilities of the Center for Computational Astrophysics, National Astronomical Observatory of Japan, including the XC-50 and XD-2000 systems. 
This work was also financially supported by JST SPRING, Grant Number JPMJSP2125. 
T. Nishikawa would like to take this opportunity to thank 
the “THERS Make New Standards Program for the Next Generation Researchers.”
\end{acknowledgments}

\appendix
\twocolumngrid
\section{On Bremsstrahlung emission from the shocked CSM}\label{sec:shockedCSM}
In the following, 
we estimate the photon number density of thermal bremsstrahlung photons emitted by the shocked CSM at the interaction point P shown in Figure~\ref{Gamma-ray_process_SNR}, 
in order to compare it with the blackbody radiation from the photosphere. 
Bremsstrahlung is expected to be the dominant radiative process in the shocked CSM, 
because the CSM heated by the SN shock reaches a temperature of the order of $T \sim 10^{9}\,{\rm K}$. 
We therefore write the bremsstrahlung emissivity in a form directly comparable to the photon number density of the blackbody radiation used in the opacity estimate of Equation~\eqref{eq:n_BB}.

The emissivity per unit volume and unit photon energy is given by
\begin{align}
  \epsilon_{\epsilon}^{\rm ff}
  &=
  \frac{2^5 \pi e^6}{3 m_{\rm e} c^3 h}
  \left( \frac{2 \pi}{3 k_{\rm B} m_{\rm e}} \right)^{1/2}
  T^{-1/2} Z_{\rm i}^2 n_{\rm e} n_{\rm i}
  \exp\!\left(-\frac{\epsilon}{k_{\rm B} T}\right)
  \bar{g}_{\rm ff}
  \notag \\
  &\left[ {\rm erg\, s^{-1}\, cm^{-3}\, erg^{-1}} \right],  
\end{align}
where $T$ is the gas temperature, $Z_{\rm i}$ is the ion charge number, and $n_{\rm i}$ and $n_{\rm e}$ are the ion and electron number densities, respectively. 
The Gaunt factor $\bar{g}_{\rm ff}$ is taken from the formulation of \citet{2011piim.book.....D}, 
their Equation~(10.5), as
\begin{align}
  \bar{g}_{\rm ff} 
  &= 
  \frac{\sqrt{3}}{\pi}
  \left[
  \ln \left(
  \frac{ \left( 2 k_{\rm B}T \right)^{3/2} }{\pi Z_{\rm i} e^2 m_{\rm e}^{1/2} \nu}
  \right)
  - \frac{5 \gamma}{2}
  \right], 
\end{align}
where $\gamma = 0.577216$ is the Euler--Mascheroni constant. 

The photon number density can then be estimated from the emissivity using a typical photon energy $\epsilon$ and the light-crossing time of the shocked CSM shell $t_{\rm lc}$ as
\begin{align}
  n_{\epsilon}^{\rm ff}
  &=
  \frac{1}{4\pi}
  \frac{\epsilon_{\epsilon}^{\rm ff}}{\epsilon} \times t_{\rm lc}. 
\end{align}
For this calculation, we consider a fixed interaction point P in the geometry shown in Figure~\ref{Gamma-ray_process_SNR}. 
For the evaluation of the point P, we adopt the geometrical parameters $\Psi_{0} = 2.39\, {\rm rad}$, $l = 5.81 \times 10^{15}\,{\rm cm}$, $d = 7.26 \times 10^{15}\,{\rm cm}$, and $\cos{\theta} = 9.88 \times 10^{-1}$. 
$R_{\rm Sh}(t)$ is taken from Model~1 of \citet{2021ApJ...922....7I}. 

We evaluate the bremsstrahlung emission at 
\begin{align}
  R_{\rm ff}(t) := \frac{R_{\rm Sh}(t)+R_{\rm CD}(t)}{2},
\end{align} 
where we denote the contact-discontinuity radius by $R_{\rm CD}(t)$, 
which corresponds to the middle of the shocked CSM shell between the shock front and the contact discontinuity. 
The light-crossing time is estimated as
\begin{align}
  t_{\rm lc}(t) = \frac{R_{\rm Sh}(t) - R_{\rm CD}(t)}{c}.
\end{align}

For the present calculation, the shocked CSM is treated as fully ionized hydrogen, so that $Z_{\rm i}=1$ and $n_{\rm e}(t) \simeq n_{\rm i}(t)$. 
The gas temperature is estimated from the shock velocity as $T = (3 m_{\rm p}/16 k_{\rm B}) V_{\rm Sh}^{2}$. 
Here we adopt $V_{\rm Sh} = 14{,}000\,{\rm km\,s^{-1}}$ from the same numerical results, which gives $T \sim 10^{9}\,{\rm K}$. 

\begin{figure}[tbp]
  \centering
  \includegraphics[width = \linewidth, height=0.6\textheight, keepaspectratio]{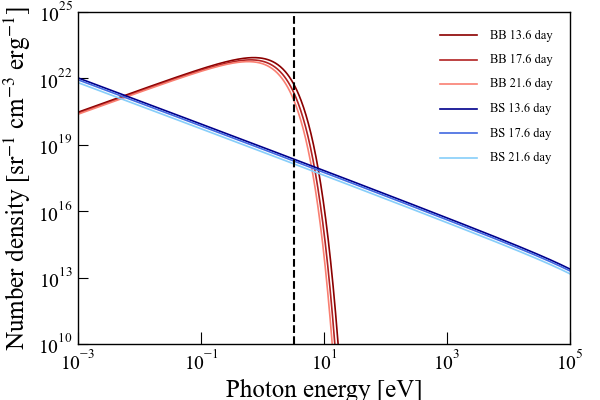} 
  \caption{
    Density spectra of soft photons from blackbody emission from the photosphere (BB) and thermal bremsstrahlung from the CSM (BS) at the interaction point P. 
    The horizontal axis represents photon energy, and the vertical axis represents photon number density. 
    The color gradient indicates the temporal evolution of the number density.  
    The black dashed line marks the energy at which the cross section for the two-photon annihilation process reaches its maximum. 
  }
  \label{fig:time-BBvsBS_2}
\end{figure}

Figure~\ref{fig:time-BBvsBS_2} shows the photon number density spectra of blackbody photons from the photosphere and bremsstrahlung photons from the shocked CSM, evaluated at the fixed interaction point P in Figure~\ref{Gamma-ray_process_SNR}. 
Under the present conditions, soft photons capable of undergoing two-photon annihilation with $E = 100\,{\rm TeV}$ gamma-rays must exceed the threshold energy, $\epsilon_{\rm min} \approx 1.69\,{\rm eV}$. 
This threshold is calculated from Equation~\eqref{eq:eps_min}. 
The maximum cross section for this process occurs when the soft photon energy is $\approx 3.31\, {\rm eV}$, as shown by the dashed line in Figure~\ref{fig:time-BBvsBS_2}, and decreases rapidly away from this value. 
From Figure~\ref{fig:time-BBvsBS_2}, it is evident that in this energy range, the number density of bremsstrahlung photons is about three orders of magnitude smaller than that of blackbody photons, indicating the dominance of the latter. 
Therefore, it is clear that 
the contribution of CSM-origin soft photons to the opacity is smaller than that of photospheric photons.

\section{CR acceleration efficiency}\label{sec:E_CR_tot}
Previous work by \citet{2020MNRAS.494.2760C} assumed a constant CR acceleration efficiency, whereas in 
\citet{2021ApJ...922....7I} 
simulations, they assumed a constant CR injection rate of $\eta_{\rm inj} = 6 \times 10^{-4}$. 
To compare the two approaches, we present the time evolution of the acceleration efficiency 
in our numerical model. 
\citet{2020MNRAS.494.2760C} assumed that the intrinsic flux is approximately proportional to the total CR energy $E_{\rm CR, tot}(t)$, and estimated its time evolution as
\begin{align}
    E_{\rm CR, tot}(t)
    &=
    \phi
    \int_{0}^{t}
    {\rm d}t'\,
    4 \pi R_{\rm Sh}^2(t')\,
    \frac{1}{2} \rho \left( R_{\rm Sh}(t') \right)
    v_{\rm Sh}^3(t'). 
\end{align}
Here, $\phi$ denotes the CR acceleration efficiency. 
If $\phi$ and the shock velocity $v_{\rm Sh}$ are constant and the ambient density decreases as $r^{-2}$, 
$E_{\rm CR, tot}(t)$ increases linearly with time. 
In this case, the intrinsic flux scales as $t^{-1}$, as shown in Equation~\ref{eq:Intr_flux_Anal_2}. 
The intrinsic flux is evaluated under these assumptions in 
\citet{2020MNRAS.494.2760C}. 

On the other hand, 
we estimate the total CR energy using the CR distribution function $f_{0}$ 
from Model~1 of \citet{2021ApJ...922....7I}: 
\begin{align}
    E_{\rm CR, tot}(t)
    &=
    \int
    {\rm d}^3 r
    \int
    pc\, f(r,p)\, {\rm d}^3 p. 
\end{align}
Since the momentum range of the distribution function calculated by \citet{2021ApJ...922....7I} spans approximately $1\,{\rm TeV}$--$10\,{\rm PeV}$, 
we extrapolate the distribution toward the lower-energy range ($1\,{\rm GeV}$--$1\,{\rm TeV}$) assuming a $p^{-4}$ dependence.

\begin{figure}[tbp]
  \centering
  \includegraphics[width = \linewidth, height=0.6\textheight, keepaspectratio]{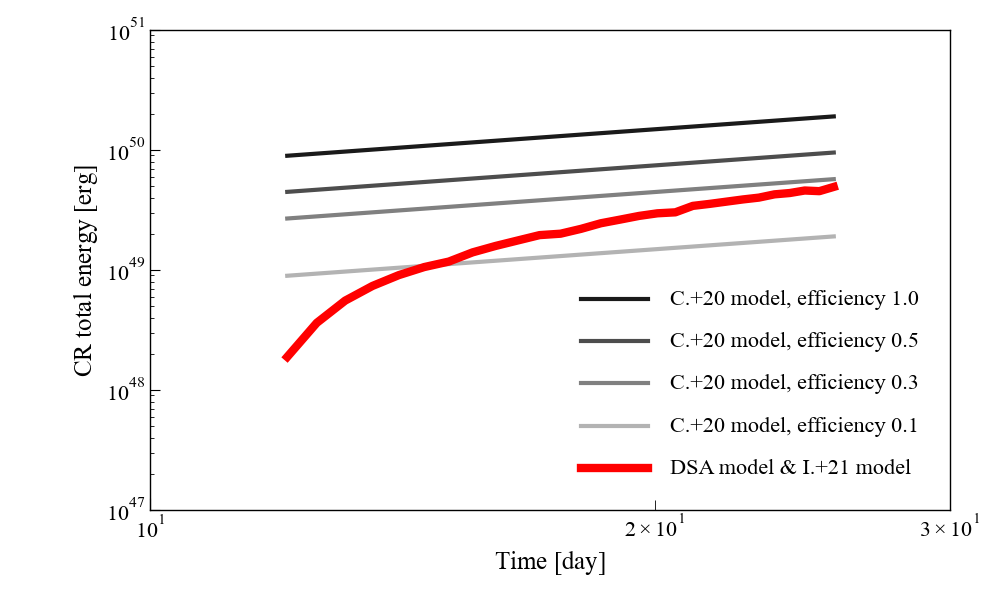} 
  \caption{
    Time evolution of the total CR energy $E_{\rm CR, tot}$ calculated using 
    the model of \citet{2020MNRAS.494.2760C} (black to gray lines) 
    and the CR distribution function from Model~1 of \citet{2021ApJ...922....7I} 
    (red curve). 
    The black-to-gray lines represent results for different acceleration efficiencies $\phi$. 
  }
  \label{fig:time-E_CR_tot_1}
\end{figure}

Figure~\ref{fig:time-E_CR_tot_1} shows the time evolution of $E_{\rm CR, tot}$ inferred from the \citet{2020MNRAS.494.2760C} model (black to gray lines) and that from \citet{2021ApJ...922....7I} (red line). 
The black-to-gray lines correspond to different values of $\phi$. 
Because \citet{2020MNRAS.494.2760C} assumes a constant acceleration efficiency $\phi$, 
the total CR energy $E_{\rm CR, tot}$ increases linearly with time.  
On the other hand, the $E_{\rm CR, tot}$ derived from \citet{2021ApJ...922....7I} 
exhibits a rise steeper than $t^{1}$ during the early phase ($t \lesssim 18\,{\rm days}$), 
and subsequently approaches a linear ($t^{1}$) trend. 
In particular, it approaches the $\phi = 0.3$ case of \citet{2020MNRAS.494.2760C}. 
As the growth of $E_{\rm CR, tot}$ becomes shallower and approaches a linear trend, 
the intrinsic flux—which is proportional to $E_{\rm CR, tot}$—is likewise expected, 
for $t \gtrsim 30\,{\rm days}$, to converge to the same $t^{-1}$ behavior as 
predicted by Equation~\ref{eq:Intr_flux_Anal_2}.

\bibliography{sample701}{}
\bibliographystyle{aasjournalv7}

\end{document}